\documentclass[12pt]{article}

\usepackage{epsf,amsfonts,hyperref}

\renewcommand{\appendix}[1]{
    \setcounter{equation}{0}
    \renewcommand{\thesection}{\Alph{section}}
    \section{Appendix: \protect\indent #1}
}
\newcommand\encadremath[1]{\vbox{\hrule\hbox{\vrule\kern8pt
\vbox{\kern8pt \hbox{$\displaystyle #1$}\kern8pt}
\kern8pt\vrule}\hrule}}
\def\enca#1{\vbox{\hrule\hbox{
\vrule\kern8pt\vbox{\kern8pt \hbox{$\displaystyle #1$}
\kern8pt} \kern8pt\vrule}\hrule}}

\newcommand\figureframex[3]{
\begin{figure}[bth]
\hrule\hbox{\vrule\kern8pt
\vbox{\kern8pt \vbox{
\begin{center}
{\mbox{\epsfxsize=#1.truecm\epsfbox{#2}}}
\end{center}
\caption{#3}
}\kern8pt}
\kern8pt\vrule}\hrule
\end{figure}
}
\newcommand\figureframey[3]{
\begin{figure}[bth]
\hrule\hbox{\vrule\kern8pt
\vbox{\kern8pt \vbox{
\begin{center}
{\mbox{\epsfysize=#1.truecm\epsfbox{#2}}}
\end{center}
\caption{#3}
}\kern8pt}
\kern8pt\vrule}\hrule
\end{figure}
}

\renewcommand{\thesection}{\arabic{section}}

\makeatletter
\@addtoreset{equation}{section}
\makeatother
\newtheorem{theorem}{Theorem}[section]
\newtheorem{conjecture}{Conjecture}[section]
\newtheorem{remark}{Remark}[section]
\newtheorem{proposition}{Proposition}[section]
\newtheorem{lemma}{Lemma}[section]
\newtheorem{corollary}{Corollary}[section]
\newtheorem{definition}{Definition}[section]
\def\br{\begin{remark}\rm\small}
\def\er{\end{remark}}
\def\bt{\begin{theorem}}
\def\et{\end{theorem}}
\def\bd{\begin{definition}}
\def\ed{\end{definition}}
\def\bp{\begin{proposition}}
\def\ep{\end{proposition}}
\def\bl{\begin{lemma}}
\def\el{\end{lemma}}
\def\bc{\begin{corollary}}
\def\ec{\end{corollary}}
\def\beaq{\begin{eqnarray}}
\def\eeaq{\end{eqnarray}}
\newcommand{\proof}[1]{{\noindent \bf proof:}\par
{#1} $\square$}

\newcommand{\eq}[1]{eq.~(\ref{#1})}

\newcommand{\beq}{\begin{equation}}
\newcommand{\eeq}{\end{equation}}
\newcommand{\bea}{\begin{eqnarray}}
\newcommand{\eea}{\end{eqnarray}}

\renewcommand{\and}{{\qquad {\rm and} \qquad}}

\newcommand{\virg}{{\qquad , \qquad}}

 \newcommand{\Tr}{{\,\rm Tr}\:}
\newcommand{\tr}{{\,\rm tr}\:}

\newcommand{\Res}{\mathop{\,\rm Res\,}}

\newcommand{\td}[1]{{\tilde{#1}}}

\renewcommand{\l}{\lambda}
\newcommand{\om}{\omega}

\newcommand{\ee}[1]{{{\rm e}^{#1}}}

\renewcommand{\d}{{{\partial}}}

\newcommand{\Pint}{{\int\kern -1.em -\kern-.25em}}

\renewcommand{\Re}{{\mathrm{Re}}}

\renewcommand{\l}{\lambda}
\renewcommand{\L}{\Lambda}

\newcommand{\bessel}{{\cal I}}

\newcommand{\Ibar}{{\hat{\cal I}}}

\newcommand{\Ybessel}{{\cal Y}}

\begin{document}
\sloppy


\pagestyle{empty}
\hfill SPhT-T08/081
\addtolength{\baselineskip}{0.20\baselineskip}
\begin{center}
\vspace{26pt}
{\large \bf {Some properties of angular integrals}}
\newline
\vspace{26pt}

{\sl M.\ Berg\`ere}\hspace*{0.05cm}\footnote{ E-mail: michel.bergere@cea.fr },
{\sl B.\ Eynard}\hspace*{0.05cm}\footnote{ E-mail: bertrand.eynard@cea.fr },
\vspace{6pt}
Institut de Physique ThŽorique,\\
CEA, IPhT, F-91191 Gif-sur-Yvette, France,\\
CNRS, URA 2306, F-91191 Gif-sur-Yvette, France.\end{center}

\vspace{20pt}
\begin{center}
{\bf Abstract}:
\end{center}

We find new representations for Itzykson-Zuber like angular integrals for arbitrary $\beta$, in particular for the orthogonal group $O(n)$, the unitary group $U(n)$ and the symplectic group $Sp(2n)$.
We rewrite the Haar measure integral, as a flat Lebesge measure integral, and we deduce some recursion formula on $n$. The same methods gives also the Shatashvili's type moments.
Finally we prove that, in agreement with Brezin and Hikami's observation, the angular integrals are linear combinations of exponentials whose coefficients are polynomials in the reduced variables $(x_i-x_j)(y_i-y_j)$.

%





\vspace{26pt}
\pagestyle{plain}
\setcounter{page}{1}



\section{Introduction}

What we call angular integral \cite{Mehtabook} is an integral over a compact Lie group $G_{\beta,n}$:
\beq
G_{1/2,n} =  O(n) \virg
G_{1,n} =  U(n) \virg
G_{2,n} =  Sp(2n) 
\eeq
of the form:
\beq
I_{\beta,n}(X,Y)=\int_{G_{\beta,n}} dO\,\, \ee{\Tr X O Y O^{-1}}
\eeq
where $X$ and $Y$ are two given matrices, and $dO$ is the Haar invariant measure on the group.
We shall also extend $I_{\beta,n}$ to arbitrary $\beta$ (Notice that our $\beta$ is half the one most commonly used in matrix models, for instance we have $\beta=1$ in the unitary case).

In this paper we are going to consider the case where $X$ and $Y$ are diagonal matrices, however, let us first recall the Harish-Chandra case.

\subsubsection*{Harish-Chandra case}

In the case where $X$ and $Y$ are in the Lie algebra of the group \cite{Knapp} (i.e. real anti-symmetric in the $O(n)$ case, anti-hermitian in the $U(n)$ case, and quaternion-anti-self-dual in the $Sp(2n)$ case), the angular integral can be computed with Weyl-character formula, and is given by the famous Harish-Chandra formula \cite{HC} (which is also a special case of the Duistermaat-Heckman localization \cite{DH}):
\beq
(X,Y)\in{\rm Lie\, algebra}\qquad\quad \Rightarrow\qquad
\int dO\,\, \ee{\Tr X O Y O^{-1}} = C\,\sum_{w\in {\rm Weyl}} {\ee{\Tr X Y_w}\over \Delta_\beta(X)\,\Delta_\beta(Y_w)}
\eeq
where $C$ is a normalization constant, $w$ runs over elements of the Weyl group, and the generalized Vandermonde determinant $\Delta_\beta(X)$ is the product of scalar products of positive roots with $X$ (see \cite{HC, PEDFZ, Knapp} for details).

\subsubsection*{Diagonal case}

However, for applications to many physics problems \cite{Dyson, Mehtabook}, it would be more interesting to have $X$ and $Y$ in other representations, and in particular {\bf $X$ and $Y$ diagonal matrices}.

Since a antihermitian matrix is, up to a multiplication by $i$, a hermitian matrix, and since every hermitian matrix can be diagonalized with a unitary conjugation, for the unitary group, the Harish-Chandra formula applies as well to the case where $X$ and $Y$ are diagonal, this is known as Itzykson-Zuber formula \cite{IZ}:
\beq
\left\{\begin{array}{l}
X={\rm diag}(x_1,\dots,x_n)\cr 
Y={\rm diag}(y_1,\dots,y_n)
\end{array}\right.
\qquad\quad \Rightarrow\qquad
\int_{U(n)} dU\,\, \ee{\Tr X U Y U^{-1}} = C_n\,{\det {\ee{x_i y_j}}\over \Delta(X)\,\Delta(Y)}
\eeq
where $\Delta(X)=\Delta_1(X)=\prod_{i>j} (x_i-x_j)$ is the usual Vandermonde determinant.

\medskip

For the other groups, computing angular integrals has remained an important challenge in mathematical physics for a rather long time. Many progresses and formulae have been found, however, a formula as compact and convenient as Harish-Chandra is still missing.
And in particular a formula which would allow to compute multiple matrix integrals, generalizing the method of Mehta \cite{Mehta1} is still missing.

\subsubsection*{Calogero Hamiltonian}

It is known that, in the diagonal case, $I_{\beta,n}$ satisfies the Calogero--Moser equation \cite{Calogero}, i.e. is an eigenfunction of the Calogero hamiltonian:
\beq
H_{\rm Calogero}\, . I_{\beta,n} = (\sum_i y_i^2)\, I_{\beta,n}
\eeq
\beq
H_{\rm Calogero} = \sum_i {\partial^2\over \partial x_i^2} + \beta \sum_{j\neq i} {1\over x_i-x_j}\,\,({\partial\over \partial x_i}-{\partial\over \partial x_j})
\eeq

Many approaches towards computing angular integrals have used that differential equation.
A basis of eigenfunctions of the Calogero hamiltonian is the Hi--Jack polynomials \cite{Calogero, BakFor, desrosiers, MD}.

In particular remarkable progress in the computation of $I_{\beta,n}$ was achieved recently by Brezin and Hikami \cite{BrezHikbeta}. By decomposing $I_{\beta,n}$ on the suitable basis of Zonal polynomials, they were able to find a recursive algorithm to compute the terms in some power series expansion of $I_{\beta,n}$, and they obtained a remarkable structure.
In particular they observed that the power series reduces to a polynomial when $\beta\in {\mathbb N}$.

\subsubsection*{Morozov and Shatashvili's formulae}

Another important question for physical applications, is not only to compute the angular integral (the partition function in statistical physics language), but also all its moments, for instance:
\beq
M_{i,j} = \int_{G_{\beta,n}} dO\,\, \ee{\Tr X O Y O^{-1}}\,\,\, ||O_{i,j}||^2
\eeq
and more generally for any indices $i_1,\dots,i_{2k},j_1,\dots,j_{2k}$:
\beq\label{shatsvilimoments}
 \int_{G_{\beta,n}} dO\,\, \ee{\Tr X O Y O^{-1}}\,\,\, O_{i_1,i_2}\,O_{i_3,i_4}\,\dots O_{i_{2k-1},i_{2k}}\,\, O^{-1}_{j_1,j_2}\,O^{-1}_{j_3,j_4}\,\dots O^{-1}_{j_{2k-1},j_{2k}}
\eeq

In the $U(n)$ case $\beta=1$, Morozov \cite{Morozov} found a beautiful formula for $M_{i,j}$, and Shatashvili \cite{Shatashvili} found a more general formula for any moments of type \ref{shatsvilimoments} using the action-angle variables of Gelfand-Tseytlin corresponding to the integrable structure of this integral.

For $\beta=1/2,1,2$, in the Harish-Chandra case where $X$ and $Y$ are in the Lie algebra, a formula for all possible moments was also derived in \cite{PEDFZ}, generalizing Morozov's \cite{eynmorozov, eynPrats}.

\medskip

In this article we shall propose new formulae for $M_{i,j}$ in the diagonal case for arbitrary $\beta$, and our method can also be generalized to all moments.

\subsubsection*{Outline of the article}

\begin{itemize}

\item Section 1 is an introduction, and we present a summary of the main results of this article.

\item In section \ref{secdefex}  we setup the notations, and we review some known examples.

\item In section \ref{secLagrange}  we show how to transform the angular integral with a Haar measure into a flat Lebesgue measure integral on a hyperplane. From it, we deduce a recursion formula, as well as a duality formula (the angular integral is an eigenfunction of kernel which is the Cauchy determinant to the power $\beta$).

\item In section \ref{secCalogero}, we discuss the moments of the angular integral. We show that moments can be obtained also with Lebesgue measure integrals, and we show that they satisfy linear Dunkl-like equations.
This can be used as a way to recover Calogero equation for the angular integral.

\item In section \ref{secpterms}, we rewrite the angular integral as a symmetric sum of exponentials with polynomial prefactors.
Those polynomials are called principal terms, and can be computed recursively.
In particular, we prove the conjecture of Brezin and Hikami \cite{BrezHikbeta} that the principal terms are polynomials in some reduced variables $(x_i-x_j)(y_i-y_j)$.

\item In section \ref{secn3}, we prove a formula for $n=3$ in terms of Bessel polynomials, and we propose a conjecture formula for arbitrary $\beta$ and arbitrary $n$.

\item In section \ref{secbeta2}, we focus on the symplectic case $\beta=2$, for which we can improve the recursion formula.

\item Section \ref{secconcl} is the conclusion.

\item Appendices contain useful lemmas, and proofs of the most technical theorems.

\end{itemize}

\subsection{Summary of the main results presented in this article}

\begin{itemize}

\item We rewrite the angular integral with the Haar measure on the Lie group $G_{\beta,n}$, as a flat Lebesgue measure integral on its Lie algebra (notations are explained in section \ref{secLagrange}):
\beq
\encadremath{
I_{\beta,n}(X;Y)
= \int dO\, \, \ee{\Tr X O Y O^{-1}}
= \int dS\,\, {\ee{\Tr S}\,\over \prod_{k=1}^n \det(S-y_k X)^{\beta}} 
}\eeq
as well as its moments:
\beq\label{introMij}
\encadremath{
\begin{array}{lll}
M_{i,j} &=& \int dO\, ||O_{i,j}||^2\,\, \ee{\Tr X O Y O^{-1}} \cr
&& \cr
&=& \beta \int dS\,\, {\ee{\Tr S}\,\over \prod_{k=1}^n \det(S-y_k X)^{\beta}} \,\,((S-y_j X)^{-1})_{i,i} 
\end{array}
}\eeq

\item We show that the $M_{i,j}$'s satisfy a linear functional equation (very similar to Dunkl operators):
\beq\label{Dunklintro}
\forall i,j,\qquad
{\partial M_{i,j}\over \partial x_i} + \beta \sum_{k\neq i} {M_{i,j}-M_{k,j}\over x_i-x_k}  = M_{i,j}\, y_j
\eeq
which implies the Calogero equation for $I_{\beta,n}=\sum_i M_{i,j}=\sum_j M_{i,j}$:
\beq
\sum_i {\partial^2 I_{\beta,n}\over \partial x_i^2} + \beta \sum_{j\neq i} {1\over x_i-x_j}\,\,({\partial I_{\beta,n}\over \partial x_i}-{\partial I_{\beta,n}\over \partial x_j})
= (\sum_i y_i^2)\, I_{\beta,n}
\eeq
Moreover, the integral of \eq{introMij}, is a solution of the linear functional equation \eq{Dunklintro} for any choice of integration domain (as long as there is no boundary term when one integrates by parts).
We thus have a large set of solutions of the linear equation, and also of Calogero equation.

\item We deduce a duality formula:
\bea
I_{\beta,n}(X;Y)
= \det(X)^{1-\beta}\,\,\int d\l_1\dots d\l_n \,\,\Delta(\L)^{2\beta}\,\, {I_{\beta,n}(X,\L)\,\over \prod_{k=1}^n\prod_{j=1}^n (\l_j-y_k)^{\beta}} 
\eea
and a recursion formula:
\beq\label{recrelintro}
\encadremath{
\begin{array}{lll}
&& I_{\beta,n}(X;Y) \cr
&=& {\ee{x_n\sum_{i=1}^n y_i}\over \prod_{i=1}^{n-1} (x_i-x_n)^{2\beta-1}}\,\,\int d\l_1,\dots d\l_{n-1}\,\, {I_{\beta,n-1}(X_{n-1},\L)\,\Delta(\L)^{2\beta}\,\,\ee{-x_n\sum_i\l_i}\over \prod_{k=1}^n\prod_{i=1}^{n-1} (\l_i-y_k)^{\beta}} \cr
\end{array}
}\eeq
similar to that of \cite{KGuhr1,KGuhr2}.

\item For $\beta\in{\mathbb N}$, the solution of the recursion can be written in terms of principal terms:
\beq
I_{\beta,n}(X,Y) = \sum_{\sigma} {\ee{\sum_{i=1}^n x_i y_{\sigma(i)}}\over \Delta(X)^{2\beta}\Delta(Y_\sigma)^{2\beta}}\,\,\, \Ibar_{\beta,n}(X,Y_\sigma)
\eeq
where $\sum_\sigma$ is the sum over all permutations.

The recursion relation \eq{recrelintro} can be rewritten as a recursion for the principal terms $\Ibar_{\beta,n}(X,Y)$:
\beq\label{recsansresintro}
\encadremath{
 \Ibar_{\beta,n}(X_n;Y_n) 
= {\Delta(Y_n)^{2\beta}\over (\beta-1)!^{n-1}}\,\, \prod_{i=1}^{n-1} x_{i,n}\,\,\,
  \left(\partial\over\partial a_i\right)^{\beta-1} \,\, {\Ibar_{\beta,n-1}(X_{n-1},a)\,\,\ee{\sum_i x_{i,n}(a_i-y_i)}\over \prod_{k=1}^n\prod_{i=1,\neq k}^{n-1} (y_k-a_i)^{\beta}} \Big|_{a_i=y_i} 
}\eeq

\item For general $n$ and $\beta$ integer, we prove the conjecture of Brezin and Hikami \cite{BrezHikbeta}, that the principal term $\Ibar_{\beta,n}(X,Y)$ is a symmetric polynomial of degree $\beta$ in the $\tau_{i,j}$ variables,
\beq
\tau_{i,j} = -\,{(x_i-x_j)(y_i-y_j)\over 2}
\eeq

\item In the case $n=3$ we find this polynomial explicitly for any $\beta$ (for $\beta$ integer the sum is finite):
\beq
\encadremath{
I_{\beta,3} 
\propto  
 \,{ \ee{x_1 y_1+x_2 y_2+x_3 y_3} \over (\Delta(x)\Delta(y))^{\beta}} \,\sum_{k=0}^{\infty} {\Gamma(\beta-k)\over 2^{6k}\, k!\,\Gamma(\beta+k)}\, \prod_{i<j}  \,\Ybessel^{(k)}_{\beta-1}({1\over \tau_{ij}})  \,\, \quad +{\rm sym}
}\eeq
where $\Ybessel_m$ is the $m^{\rm th}$ Bessel polynomial, i.e. the modified Bessel function of the second kind (see definition of $\Ybessel_{\beta-1}$ below in \eq{DefYbessel}).

\item In the case $\beta=2$ (i.e. symplectic group $Sp(2n)$), the recursion relation for the principal term can be written:
\beq
\begin{array}{lll}
\Ibar_{2,n} 
&=&  \prod_{i=1}^{n-1} (x_i-x_n)\, (y_i-y_n)^2\,\cr
&& \prod_{i=1}^{n-1} \Big(
x_i-x_n - \sum_{k=1, \neq i}^{n} {2\over y_i-y_k} + {\partial\over\partial a_i}
\Big)
\Ibar_{2,n-1}(X_{n-1},a)\,\, \Big|_{a_i=y_i} \cr
&=& \Delta(X_n)^2\Delta(Y_n)^2\,\,\,   {\det\left[ X_{n-1}-x_{n}-{2\over Y_{n-1}-y_{n}}+B+\partial_Y \right]\over \det(X_n-x_{n})}
 \,\,{\hat{I}_{2,n-1}(X_{n-1};Y_{n-1}) \over \Delta(X_{n-1})^2\Delta(Y_{n-1})^2}
\end{array}
\eeq
and $B$ is the antisymmetric matrix $B_{i,j}={\sqrt{2}\over y_i-y_j}$, $B_{i,i}=0$, and $\partial_Y={\rm diag}(\partial_{y_1},\dots,\partial_{y_{n-1}})$.
In section \ref{secbeta2} we propose an operator formalism to compute it, and we propose a conjecture formula in terms of decomposition into triangles.

\end{itemize}

\section{Definitions and examples}
\section{secdefex}

\subsection{Notations for angular integrals}

Let $X$ and $Y$ be two diagonal matrices of size $n$:
\beq
X={\rm diag}(x_1,\dots,x_n)
\virg
Y={\rm diag}(y_1,\dots,y_n)
\eeq
We define the {\bf angular integral}:
\beq
I_{\beta,n}(x_1,\dots,x_n;y_1,\dots,y_n) = \int_{G_{\beta,n}}\,\, dO \,\,\ee{\Tr X O Y O^{-1}}
\eeq
where $G_{\beta,n}$ denotes one of the Lie groups:
\beq
G_{1/2,n} =  O(n) 
\virg
G_{1,n} =  U(n) 
\virg
G_{2,n} =  Sp(2n) 
\eeq
and $dO$ is the invariant Haar measure on the corresponding compact Lie group.

We will later extend those notions to arbitrary values of $\beta$.

\bigskip


\subsection{Bessel polynomials}

For further use, we need to introduce some Bessel functions \cite{Abm, Krall, WikiBessel, WikiBesselPol, Carlitz}. Those special functions are going to play a major role throughough this article.
\smallskip


The Bessel polynomials (see \cite{Krall, WikiBesselPol}) $\Ybessel_m(x)$ are defined by:
\beq\label{DefYbessel}
\Ybessel_m(x) = \sum_{k=0}^\infty {\Gamma(m+k+1)\over k!\,\Gamma(m-k+1)}\,(x/2)^k
=\sqrt{2 \over \pi x}\,\ee{1/x}\,{\cal K}_{m+{1\over 2}}(1/x)
\eeq
where ${\cal K}$ is the modified Bessel function of the second kind \cite{Abm, WikiBessel}.
$\Ybessel_m$ is a polynomial of degree $m$ when $m$ is an integer:
\beq
\Ybessel_0 = 1  \, , \,\,\,\,
\Ybessel_1 = x+1  \, , \,\,\,\,
\Ybessel_2 = 3x^2+3x+1  \, , \,\,\,\,
\Ybessel_3 = 15 x^3+ 15 x^2+6x+1  \, , \,\,\,
{\rm etc}\dots
\eeq
They satisfy:
\beq
x^2 \Ybessel_m'' +(2x+2)\Ybessel_m' -m (m+1)\Ybessel_m =0
\eeq

\bigskip
We shall also need:
\beq
Q_{\beta,j}(x) = \sum_{k=0}^{\infty}\, {\Gamma(\beta+j+k)\over k!\, \Gamma(\beta-j-k)}\,2^{-k}\, x^{\beta-j-k}
\eeq
which is a polynomial of degree $\beta-j$ if $\beta$ is an integer.

In particular $Q_{\beta,0}$ is the Carlitz polynomial \cite{Carlitz, WikiBesselPol} and is closely related to $\Ybessel_{\beta-1}$:
\beq
Q_{\beta,0}(x) = x^\beta\,\Ybessel_{\beta-1}({1\over x})  =  \sqrt{2\over \pi}\,\,\ee{x}\,\,x^{\beta+{1\over 2}}\,\, K_{\beta-{1\over 2}}(x)
\eeq
satisfying:
\beq
x^2 Q_{\beta,0}'' - 2x(\beta+x) Q_{\beta,0}' + 2\beta (x+1) Q_{\beta,0}=0
\eeq
The first fews are:
\beq
Q_{1,0} = x  \, , \,\,\,\,
Q_{2,0} = x^2+x  \, , \,\,\,\,
Q_{3,0} = x^3+3x^2+3x  \, , \,\,\,\,
Q_{4,0} = x^4 + 6 x^3 + 15 x^2+ 15 x  \, , \,\,\,
{\rm etc}\dots
\eeq

For higher $j$, the $Q_{\beta,j}$'s are derivatives of Bessel polynomials:
\beq
Q_{\beta,j}(1/x) 
= 2^{j}\,x^{j-\beta}\, {d^j \over dx^j} \,\Ybessel_{\beta-1}(x) 
= 2^{j}\,x^{j-\beta}\, \,\Ybessel^{(j)}_{\beta-1}(x) 
\eeq
They satisfy:
\beq
-x\, Q_{\beta,j}
={1\over 4}\,   Q_{\beta,j+1}  + j  Q_{\beta,j}  
+ (j-\beta)(j+\beta-1) Q_{\beta,j-1} 
\eeq
\beq
Q_{\beta,j+1} 
= 2 (\beta-j-x{d\over dx}) Q_{\beta,j} 
\eeq
The first fews are:
\beq
Q_{2,1} = 2x  \, , \,\,\,\,
Q_{3,1} = 6 x^2+12 x  \, , \,\,\,\,
Q_{4,1} = 12 x^3+60 x^2 + 90 x  \, , \,\,\,
\eeq
\beq
Q_{3,2} = 24 x  \, , \,\,\,\,
Q_{4,2} = 120 x^2 + 360 x  \, , \,\,\,\,
Q_{4,3} = 720 x  \, , \,\,\,
{\rm etc}\dots
\eeq


\subsection{Examples angular integrals with $n=1,2,3$}

{\bf $\bullet$ $n=1$:} The $n=1$ case needs no computation, and gives:
\beq
I_{\beta,1}(x;y) = \ee{xy}
\eeq

\noindent {\bf $\bullet$ $n=2$:} The $n=2$ case requires a little bit of easy computation, and it has been known for some time, we have (this formula is rederived in this article):
\bea
I_{\beta,2}(X,Y) 
&=& {\ee{x_1 y_1 + x_2 y_2}\over \tau^\beta}\,\,\Ybessel_{\beta-1}(1/\tau) + {\ee{x_1 y_2 + x_2 y_1}\over (-\tau)^\beta}\,\,\Ybessel_{\beta-1}(-1/\tau) \cr
&=& {\ee{x_1 y_1 + x_2 y_2}\over \tau^{2\beta}}\,\,Q_{\beta,0}(\tau) + {\ee{x_1 y_2 + x_2 y_1}\over (-\tau)^{2\beta}}\,\,Q_{\beta,0}(-\tau) 
\eea
where
\beq
\tau = - {1\over 2}(x_1-x_2)(y_1-y_2) 
\eeq

It can also be written in terms of the modified Bessel function $\bessel$:
\beq
I_{\beta,2}(X;Y) 
= {\ee{{1\over 2}(x_1+x_2)(y_1+y_2)}\over \tau^{2\beta-1}}\, \bessel_{\beta-{1\over 2}}(\tau)  \virg
\eeq
where
\beq
\bessel_m(\tau) =(\tau/2)^{2m} \sum_{k=0}^\infty {(\tau/2)^{2k}\over k! \Gamma(m+k+1)}
\virg \bessel_m = \bessel_m'' + {1-2m\over \tau} \bessel_m' 
\eeq


\bigskip
\noindent {\bf $\bullet$ $n=3$:} 

we show in this article that (proof in appendix \ref{appproofn3}):
\beq
\encadremath{
I_{\beta,3} 
\propto  
 \,{ \ee{x_1 y_1+x_2 y_2+x_3 y_3} \over (\Delta(x)\Delta(y))^{\beta}} \,\sum_{k=0}^{\infty} {\Gamma(\beta-k)\over 2^{6k}\, k!\,\Gamma(\beta+k)}\, \prod_{i<j}  \,\Ybessel^{(k)}_{\beta-1}({1\over \tau_{ij}})  \,\, \quad +{\rm perm.}
}\eeq
where
\beq
\tau_{i,j} = -\,{(x_i-x_j)(y_i-y_j)\over 2}
\eeq
and $+{\rm perm.}$ means that we have to symmetrize over all permutations of the $y_j$'s.


\bigskip
\noindent {\bf $\bullet$ $n> 3$:} 
We show in this article that for arbitrary $n$ and $\beta$, the angular integral is of the form conjectured by Brezin and Hikami:
\beq
I_{\beta,n} 
\propto  
 \,{ \ee{\sum_i x_i y_i} \over (\Delta(x)\Delta(y))^{2\beta}} \Ibar_{\beta,n}(\tau_{ij})  \,\, \quad +{\rm perm.}
\eeq
where $\Ibar_{\beta,n}(\tau_{ij})$ is a polynomial in the $\tau_{i,j}$'s, and for which we write  a recursion relation.

\section{Transformation of the angular integral}\label{secLagrange}

In this section, we transform the Haar measure group integral into a flat Lebesgue measure integral.

\subsection{Lagrange multipliers}

For $\beta=1/2,1,2$, an element  $O\in G_{\beta,n}$ is an orthonormal basis, i.e. a collection of $n$ orthonormal vectors $e_1,\dots, e_n$, whose coordinates $O_{i,j}=(e_i)_j$ are of the form:
\beq
(e_i)_j=O_{i,j} = \sum_{\alpha=0}^{2\beta-1} (e_i)^\alpha_j\,\, \epsilon_\alpha
\eeq
where the $\epsilon_\alpha$'s form a basis of a Clifford algebra (indeed this reproduces the three groups $G_{\beta,n}$ for $\beta=1/2,1,2$):
\beq
\epsilon_0=1
\,\,, \,\, \epsilon_0^\dagger = \epsilon_0 \virg
\forall\, \alpha>0:\quad
\epsilon_\alpha^2=-1
\,\,\,\,, \,\,\, \epsilon_\alpha^\dagger = -\epsilon_\alpha 
\virg \epsilon_\alpha.\epsilon_{\alpha'}=-\epsilon_{\alpha'}.\epsilon_\alpha
\eeq
with structure constants (only for $\beta=2$):
\beq
\epsilon_\alpha \epsilon_{\alpha'}^\dagger = \sum_{\alpha''} \eta_{\alpha,\alpha',\alpha''}\, \epsilon_{\alpha''}
\eeq
and where $\eta_{\alpha,\alpha',\alpha''}$ has
the property useful for our purpose, that for every pair $(\alpha,\alpha')$, there is exactly only one $\alpha''$ such that $\eta_{\alpha,\alpha',\alpha''}\neq 0$.
In particular $\eta_{\alpha,\alpha,\alpha''}=\delta_{\alpha'',0}$.


\medskip
The basis must be orthonormal, i.e.
\beq
e_i.e_j^\dagger = \delta_{i,j} = \sum_{k=1}^n (e_i)_k \, (e_j)^\dagger_k  = \sum_{k=1}^n\sum_{\alpha,\alpha'=0}^{2\beta-1} (e_i)^\alpha_k \, (e_j)^{\alpha'}_k \,\, \epsilon_\alpha\, \epsilon_{\alpha'}^\dagger
\eeq
We introduce Lagrange multipliers to enforce those orthonormality relations
\beq
\delta(e_i.e_i^\dagger-1) = \int dS_{i,i}\,\,\, \ee{S_{i,i}(1-\sum_{k,\alpha} ((e_i)^\alpha_k)^2}
\eeq
and if $i<j$:
\bea
\delta(e_i.e_j^\dagger) 
&=& \int \dots \int dS_{i,j}^0,\dots dS_{i,j}^{2\beta-1}\,\,\, \, \ee{ - 2\sum_{\alpha,\alpha',\alpha''}\sum_k S_{i,j}^\alpha ((e_i)^{\alpha'}_k)((e_j)^{\alpha''}_k) \eta_{\alpha',\alpha'',\alpha}} \cr
&=&  \int dS_{i,j}\,\,\, \, \ee{ - 2 \sum_k S_{i,j} ((e_i)_k)((e_j)^\dagger_k) } 
\eea
where each integral is over the imaginary axis.
\medskip

Since the scalar product is invariant under group transformations (i.e. change of orthogonal basis), the following measure is invariant and thus must be proportional to the Haar measure:
\beq
dO \propto \prod_{i,j,\alpha} d (e_i)^\alpha_j\,\, \prod_i \delta(e_i.e_i^\dagger-1)\prod_{i<j} \delta(e_i.e_j^\dagger)
\eeq 
i.e.
\beq\label{Haarmeasureden}
dO \propto \prod_{i,j,\alpha} d (e_i)^\alpha_j\,\, \int dS\,\, 
\ee{\sum_i S_{i,i}}
\,\, \ee{ -\sum_i  \sum_{k}  S_{i,i} |(e_i)_k|^2}  
 \,\,\ee{ - 2\sum_{i<j} \sum_k S_{i,j}^\alpha \,\, (e_i)_k\, (e_j)^\dagger_k }
\eeq
where
\beq
dS = \prod_i dS_{i,i} \prod_{i<j} \,\, dS_{i,j} = \prod_{i=1}^n dS_{i,i}\,\, \prod_{i<j} \prod_{\alpha=0}^{2\beta-1} dS_{i,j}^\alpha
\eeq
is the $G_{\beta,n}$ invariant measure on the space $E_{\beta,n}$:
\beq
iS\in \left\{\begin{array}{l}
E_{1/2,n}=\{n\times n\,\, {\rm real\, symmetric\, matrices}\} \cr
E_{1,n}=\{n\times n\,\, {\rm hermitian\, matrices}\} \cr
E_{2,n}=\{n\times n\,\, {\rm quaternion\, self-dual\, matrices}\} \cr
\end{array}\right.
\eeq
where we have completed $S$ by self duality ($S=S^\dagger$):
\beq
S_{j,i}^0 = S_{i,j}^0
\quad , \,\, {\rm and}\, \forall\, \alpha>0\quad
S_{j,i}^\alpha = -S_{i,j}^\alpha
\eeq

Therefore we have (up to a multiplicative constant):
\bea
I_{\beta,n}(X;Y)
&\propto& \int dS\,\int de_1\,\dots\, de_n \,\, \ee{\sum_i S_{i,i}}
\,\,\ee{\sum_{i,k} x_i y_k |(e_i)_k|^2}\, \cr
&& \,\, \ee{ -\sum_i  \sum_{k} S_{i,i} |(e_i)_k|^2}  
 \,\,\ee{ - 2\sum_{i<j} \sum_k \sum_{\alpha,\alpha',\alpha''} S_{i,j}^\alpha ((e_i)^{\alpha'}_k)({(e_j)^\dagger}^{\alpha''}_k)\,\eta_{\alpha',\alpha'',\alpha} } \cr
\eea
The integral over the $(e_i)^\alpha_k$'s is now gaussian and can be performed.
The gaussian integrals for each $k$ are independent.

The quadratic form in the exponential is, for each $k$:
\beq
\sum_{\alpha,\alpha',\alpha''} \sum_{i,j}\, (e_i)_k^\alpha (e_j)_k^{\alpha'} \eta_{\alpha,\alpha',\alpha''}
(\delta_{i,j} x_i y_k \delta_{\alpha'',0} - S_{i,j}^{\alpha''})
\eeq
If we define the vector $v_k=(v_{1,k},\dots,v_{n,k})$ where $v_{i,k} = \sum_\alpha (e_i)_k^\alpha \epsilon_\alpha^\dagger$, we have to compute the gaussian integral:
\beq
\int dv_k \,\, \ee{- v_k^\dagger (S-y_k X ) v_k}
\eeq
For the 3 values of $\beta=1/2,1,2$, this integral is worth:
\beq
\int dv_k \,\, \ee{- v_k^\dagger (S-y_k X ) v_k}
= {(2\pi)^\beta\over \det(S-y_k X)^\beta}
\eeq
where $\det$ is the product of singular values (see \cite{Mehtabook}).

Thus we get the following theorem:

\bt The angular integral $I_{\beta,n}(X;Y)$ is also equal to the following flat Lebesgue measure integral:
\beq
\encadremath{
I_{\beta,n}(X;Y)
\propto \int dS\,\, {\ee{\Tr S}\,\over \prod_{k=1}^n \det(S-y_k X)^{\beta}} 
= \det(X)^{1-\beta}\,\,\int dS\,\, {\ee{\Tr S X}\,\over \prod_{k=1}^n \det(S-y_k)^{\beta}} 
}\eeq

\et
In the last formula we have made the change of variable $S\to X^{1/2} S X^{1/2}$.
Also, the integration domain for $S$, which was $i E_{\beta,n}$ before exchanging the integrations over $S$ and $e$, is now shifted to the right, so that all singular values of $(S-y_k X)$ have positive real part.
The integration domain for $S$ can be deformed such that the integral remains convergent and the integration path goes to the right of all zeroes of the denominator.
If $\beta$ is half-integer or integer, the denominator is not singular near $\infty$, and the integration contour can be closed.
This will be made more precise below.

\medskip
{\bf Remark 1:} For the moment, this formula holds only for $\beta=1/2,1,2$. Later we will extend it to other values of $\beta$.

\medskip
{\bf Remark 2:} 
Another remark, is that a similar formula can be obtained by exchanging the roles of $X$ and $Y$.

\subsection{Duality formula}

Notice that the matrix $S$ itself can be diagonalized with a $G_{\beta,n}$ conjugation:
\beq
S = O \L O^{-1}
\virg
\L={\rm diag}(\l_1,\dots,\l_n)
\,\,\, ,
\,\,\,
O\in G_{\beta,n}
\eeq
and the measure $dS$ is up to a constant \cite{Mehtabook}:
\beq
dS \propto dO\,\, d\L\,\,\, \Delta(\L)^{2\beta}
\eeq
Therefore, the angular integral reappears in the RHS:
\bea
I_{\beta,n}(X;Y)
&\propto& \det(X)^{1-\beta}\,\,\int dS\,\, {\ee{\Tr S X}\,\over \prod_{k=1}^n \det(S-y_k)^{\beta}} \cr
&\propto& \det(X)^{1-\beta}\,\,\int d\L \,\,\Delta(\L)^{2\beta}\,\, {I_{\beta,n}(X,\L)\,\over \prod_{k=1}^n\prod_{j=1}^n (\l_j-y_k)^{\beta}} \cr
\eea
Here, if we assume that $\forall i,\, x_i\in {\mathbb R}^+$, the integration contours for the $\l_i$'s are of the form $r+i{\mathbb R}$ where $r> {\rm max}(\Re \,y_k)$. If $\beta$ is integer or half integer, the denominator in the integrand is not singular near $\infty$, and the integration contour can be closed.
Thus, if $2\beta$ is an integer, the integration contours for the $\l_i$'s can be choosen as circles of radius $>{\rm max}(|y_k|)$.

This equation looks better if we rewrite it in term of  the Cauchy determinant $D_n(X,Y)$:
\beq\label{defDnCauchy}
D_n(X,Y) = \det\left(1\over x_i-y_j\right) = {\Delta(X)\Delta(Y)\over \prod_{i,j}(x_i-y_j)}
\eeq
and the rescaled function
\beq\label{defcheckI}
\check{I}_{\beta,n}(x_1,\dots,x_n;y_1,\dots,y_n) = 
(\Delta(X)\Delta(Y))^{\beta}\,\,I_{\beta,n}(x_1,\dots,x_n;y_1,\dots,y_n) 
\eeq
We then have:

\bt  the rescaled function $\check{I}_{\beta,n}$ satisfies the duality formula:
\beq
\encadremath{
\check{I}_{\beta,n}(X;Y)
\propto \det(X)^{1-\beta}\,\,\int d\L\, \check{I}_{\beta,n}(X,\L)\, \,D_n(\L,Y)^{\beta} 
}\eeq
i.e. $\check{I}_{\beta,n}(X;Y)$ is an eigenfunction of the kernel $D_n^\beta$. 
\et

\bigskip
\noindent {\bf Remark 1:} The duality formula above was derived for $\beta=1/2,1,2$, but it makes sense for any $\beta$.

\bigskip
\noindent {\bf Remark 2:} It is easy to check that this relation is satisfied for the Itzykson-Zuber case $\beta=1$, indeed in that case we have $\check{I}_{1,n}(X;Y)=\det(\ee{x_i y_j}) = \sum_\rho (-1)^\rho \,\prod_i \ee{y_i x_{\rho(i)}}$, and:
\bea
&& \int d\L\, \check{I}_{1,n}(X,\L)\, \,D_n(\L,Y) \cr
&\propto& \sum_{\sigma,\rho} (-1)^\sigma\, (-1)^\rho\,\,  \int \prod_{i=1}^n \, {\ee{\l_i\,x_{\rho(i)}}\over \l_i-y_{\sigma(i)}}\,\, d\l_i \cr
&=& \sum_{\sigma,\rho} (-1)^\sigma\, (-1)^\rho\,\,  \prod_{i=1}^n \, \ee{y_{\sigma(i)}\,x_{\rho(i)}} \cr
&=& n!\, \det(\ee{x_i y_j}) \cr
&\propto& \check{I}_{1,n}(X,Y)
\eea

\subsection{Recursion formula}

First, let us notice that we can always assume that $x_n=0$, otherwise we perform a shift $X\to X-{x_n}$:
\bea
I_{\beta,n}(X,Y) 
&=& \int_{G_{\beta,n}}\,\, dO\,\, \ee{\Tr X O Y O^{-1}} \cr
&=& \int_{G_{\beta,n}}\,\, dO\,\, \ee{\Tr (X-x_n) O Y O^{-1}}\,\,\ee{x_n \Tr  O Y O^{-1}} \cr
&=& \ee{x_n\Tr Y}\,\,\int_{G_{\beta,n}}\,\, dO\,\, \ee{\Tr (X-x_n) O Y O^{-1}}
\eea
Thus we define:
\beq
X_{n-1}={\rm diag}(x_1,\dots,x_{n-1})
\virg
\td{X} = X_{n-1}-x_n\,{\rm Id}_{n-1}
\eeq

\bigskip

Then, we notice that the orthonormality of the basis $e_i$:
\beq
e_i.e_j^\dagger =\delta_{i,j}
\eeq
implies that if we already know $e_1,\dots,e_{n-1}$, then $e_n$ is completely fixed (up to an irrelevant phase).
In other words, it is sufficient to enforce only the orthonormality of $e_1,\dots,e_{n-1}$ with Lagrange multipliers, i.e. introduce a matrix $S$ of size $n-1$.

Also, because of our shift $X\to X-x_n$, we notice that $e_n$ does not appear in the integrand.

Then, we write as in eq.\ref{Haarmeasureden}:
\bea\label{Haarmeasuredenm1}
dO 
&\propto& \prod_{i=1}^{n-1} d e_i\,\, \prod_{i=1}^{n-1} \delta(e_i.e_i-1)\prod_{i<j=1}^{n-1} \delta(e_i.e_j^\dagger) \cr
&\propto& \prod_{i=1}^{n-1}\prod_{j=1}^{n} \prod_{\alpha=0}^{2\beta-1} d (e_i)^\alpha_j\,\, \int_{i E_{\beta,n-1}} dS\,\, 
\ee{\sum_i S_{i,i}}
\,\, \ee{ -\sum_i  \sum_{k} S_{i,i} |(e_i)_k|^2}  
 \,\,\ee{ - 2\sum_{i<j} \sum_k S_{i,j}\, (e_i)_k\,(e_j)^\dagger_k }\cr
&\propto& \prod_{i=1}^{n-1} d e_i\,\, \int_{i E_{\beta,n-1}} dS\,\, 
\ee{\Tr S}
\,\, \ee{ - \sum_{i,j} S_{i,j} \,\,e_i. e_j^\dagger}
 \eea
which implies, after performing the gaussian integral over the $e_1,\dots,e_{n-1}$:
\bea
I_{\beta,n}(X;Y)
&\propto& \ee{x_n\tr Y}\,\,\int_{i E_{\beta,n-1}} dS\,\, {\ee{\Tr S}\,\over \prod_{k=1}^n \det(S-y_k \td{X})^{\beta}} \cr
&\propto& {\ee{x_n\tr Y}\over \prod_{i=1}^{n-1} (x_i-x_n)^{2\beta-1}}\,\,\int_{i E_{\beta,n-1}} dS\,\, {\ee{\Tr S\td{X}}\,\over \prod_{k=1}^n \det(S-y_k)^{\beta}} \cr
\eea
Again, $S$ can be diagonalized:
\beq
S = O \L O^{-1}
\virg
\L={\rm diag}(\l_1,\dots,\l_{n-1})
\,\,\, ,
\,\,\,
O\in G_{\beta,n-1}
\eeq
i.e. the rank $n$ angular integral $I_{\beta,n}$ is expressed in terms of the rank $n-1$:
\bea
I_{\beta,n}(X;Y)
&\propto& {\ee{x_n\tr Y}\over \prod_{i=1}^{n-1} (x_i-x_n)^{2\beta-1}}\,\,\int d\L\,\, {I_{\beta,n-1}(\td{X},\L)\,\Delta(\L)^{2\beta}\over \prod_{k=1}^n\prod_{i=1}^{n-1} (\l_i-y_k)^{\beta}} \cr
&\propto& {\ee{x_n\tr Y}\over \prod_{i=1}^{n-1} (x_i-x_n)^{2\beta-1}}\,\,\int d\L\,\, {I_{\beta,n-1}(X_{n-1},\L)\,\Delta(\L)^{2\beta}\,\,\ee{-x_n\sum_i\l_i}\over \prod_{k=1}^n\prod_{i=1}^{n-1} (\l_i-y_k)^{\beta}} \cr
\eea
Which is our main recursion formula:

\bt\label{thmainrecIbetan}
The angular integrals $I_{\beta,n}(X;Y)$ satisfy the recursion:
\beq\label{mainrecIbetan}
\encadremath{
\begin{array}{lll}
&& I_{\beta,n}(X;Y) \cr
&& \cr
&\propto& {\ee{x_n\sum_{i=1}^n y_i}\over \prod_{i=1}^{n-1} (x_i-x_n)^{2\beta-1}}\,\,\int d\l_1,\dots d\l_{n-1}\,\, {I_{\beta,n-1}(X_{n-1},\L)\,\Delta(\L)^{2\beta}\,\,\ee{-x_n\sum_i\l_i}\over \prod_{k=1}^n\prod_{i=1}^{n-1} (\l_i-y_k)^{\beta}} 
\end{array}
}\eeq
\et
Here again, the integration contours for the $\l_i$'s are such that the integral is convergent, and such that they surround all the $y_k$'s.
For instance, if $2\beta$ is an integer, and if $\forall i\, x_i\in {\mathbb R}^+$, the integration contour for the $\l_i$'s can be choosen as circles of radius $>{\rm max}(|y_k|)$.

\br
Now this recursion formula can be used to define $I_{\beta,n}$ for arbitrary $\beta$, so that it coincides with the angular integral for $\beta=1/2,1,2$.
\er


We have also the iterated form:
\bea
I_{\beta,n}(X;Y)
&\propto& {\ee{x_n\sum_{i=1}^n y_i}\over \Delta(X)^{2\beta-1}}\,\, \int \prod_{i=1}^{n-1}\prod_{j=1}^{i} d\l_{i,j} \cr
&& \,\, {\prod_{i=1}^{n-1} \prod_{1\leq j<j'\leq i} (\l_{i,j'}-\l_{i,j})^{2\beta}\,\,\prod_{i=1}^{n-1} \ee{(x_{i}-x_{i+1}) \sum_j \l_{i,j}}\over \prod_{i=1}^{n-1}\prod_{j=1}^{i+1}\prod_{j'=1}^i (\l_{i,j'}-\l_{i+1,j})^{\beta}} \cr
\eea
where we have defined $\l_{n,j}=y_j$, and where the integration contours are circles such that:
\beq
|\l_{i,j}| = \rho_i \virg \rho_1 > \rho_2  >\dots > \rho_{n-1} > {\rm max} |y_j| 
\eeq

\bigskip

\br
A similar recursion relation was also found Kohler and Guhr \cite{KGuhr1,KGuhr2,KGuhr3}, but the authors found real integrals instead of contour integrals. The advantage of our formulation, is that we can easily move integration contours and find new relations, as we will see below.
\er


\section{Moments of angular integrals and Calogero}
\label{secCalogero}

In this section, we compute moments of the angular integral, and we show that our formula indeed satisfies Calogero equation.

\subsection{Generalized Morozov's formula}

Define the quadratic moments (see \cite{Morozov} for $\beta=1$):
\beq\label{defMorozovmoments}
M_{i,j} = \int_{G_{\beta,n}}\,\, dO \,\, ||O_{i,j}||^2\,\,\ee{\Tr X O Y O^{-1}}
\eeq
The same calculation as above yields:
\bea\label{MijintdS}
M_{i,j} 
&=& \beta \int dS\,\, {\ee{\Tr S}\,\over \prod_{k=1}^n \det(S-y_k X)^{\beta}} \,\,((S-y_j X)^{-1})_{i,i} \cr
&=& {\beta\over x_i}\,\det(X)^{1-\beta}\, \int dS\,\, {\ee{\Tr SX}\,\over \prod_{k=1}^n \det(S-y_k )^{\beta}} \,\,((S-y_j )^{-1})_{i,i} \cr
\eea

\bigskip

As a consistency check, and as a warmup exercise, let us show that this formula satisfies:
\beq\label{IsumMij}
I_{\beta,n} =  \sum_j M_{i,j}
\eeq
which comes from $\forall O\in G_{\beta,n},\,\,\, \sum_j ||O_{i,j}||^2 = 1$.

\bigskip

We have:
\bea
\sum_j M_{i,j} 
&=& \sum_j {\beta\over x_i}\,\det(X)^{1-\beta}\, \int dS\,\, {\ee{\Tr SX}\,\over \prod_{k=1}^n \det(S-y_k )^{\beta}} \,\,((S-y_j )^{-1})_{i,i} \cr
&=& {-1\over x_i}\,\det(X)^{1-\beta}\, \int dS\,\, \ee{\Tr SX}\,\,{\d\over \d S_{i,i}}\,  {1\over \prod_{k=1}^n \det(S-y_k )^{\beta}}  \cr
&=& {1\over x_i}\,\det(X)^{1-\beta}\, \int dS\,\, {1\over \prod_{k=1}^n \det(S-y_k )^{\beta}}\,\, {\d\over \d S_{i,i}}\,\ee{\Tr SX}   \cr
&=& \det(X)^{1-\beta}\, \int dS\,\, {1\over \prod_{k=1}^n \det(S-y_k )^{\beta}}\,\, \ee{\Tr SX}   \cr
&=& I_{\beta,n}
\eea
Notice that this equality holds independently of the integration domain of $S$, provided that one can integrate by parts without picking boundary terms.

\medskip
{\bf Remark:}
Of course a similar equation can be found by exchanging the roles of $X$ and $Y$, and one gets
symmetrically:
\beq\label{IsumMijXtoY}
\sum_i M_{i,j} = I_{\beta,n}
\eeq

\subsection{Other moments}

Since, after introducing the Lagrange multipliers, the integral becomes gaussian in the $O_{i,j}$'s, any polynomial moment can be computed using Wick's theorem. It is sufficient to compute the propagator:
\beq
< O_{i,k}\, O_{j,l}^\dagger> = \beta\, \delta_{k,l}\, \left((S-y_k\, X)^{-1}\right)_{i,j}
\eeq
Then, the expectation value of any polynomial moment is obtained as the sum over all pairings of  the product of propagators.

For instance:
\bea
&& \int dO\,\, \ee{\Tr X O Y O^{-1}}\,\,\,\, O_{i_1,j_1}\, O_{i_2,j_2}\,O_{i_3,j_3}^\dagger\,O_{i_4,j_4}^\dagger \cr
&=& \int dS\, {\ee{\Tr S}\,\over \prod_{k=1}^n \det(S-y_k X)^{\beta}} \,\,
\Big[ \cr
&& 
\delta_{j_1,j_3}\, \delta_{j_2,j_4}\,
\left((S-y_{j_1}\, X)^{-1}\right)_{i_1,i_3} \left((S-y_{j_2}\, X)^{-1}\right)_{i_2,i_4} \cr
&& +
\delta_{j_1,j_4}\, \delta_{j_2,j_3}\, 
\left((S-y_{j_1}\, X)^{-1}\right)_{i_1,i_4} \left((S-y_{j_2}\, X)^{-1}\right)_{i_2,i_3}
\Big]\cr
\eea
In principle, one could compute with this method the generalization of all Shatashvili's moments \cite{Shatashvili}.

\subsection{Linear equations}

In this section we prove that the $M_{i,j}$'s satisfy the following linear functional relations, which are very similar to Dunkl equations \cite{}:
\beq\label{eqdiflinMij}
\encadremath{
\forall i,j \virg
{\partial M_{i,j}\over \partial y_j} + \beta \sum_{l\neq j} {M_{i,l}-M_{i,j}\over y_l-y_j}  = M_{i,j}\, x_i
}\eeq

We are going to give 2 different proofs of \eq{eqdiflinMij}. The first one below is based on integration by parts. It can be done for the 3 groups $\beta=1/2,1,2$, however it is rather tedious for $\beta=1/2$ and $\beta=2$, and we present the proof only for $\beta=1$.
Another proof valid for all 3 values of $\beta$ is presented in section \ref{secproofDunklloop} below.

\medskip
Let us check that \eq{MijintdS} satisfies \eq{eqdiflinMij} (for $\beta=1$).
We first rewrite:
\bea
{1\over y_l-y_j}\,\left({1\over S-y_l} - {1\over  S-y_j}\right)_{i,i}  
&=& ((S-y_l)^{-1}(S-y_j)^{-1})_{i,i}  \cr
&=& \sum_m  ((S-y_l)^{-1})_{i,m}\,\,\, ((S-y_j)^{-1})_{m,i}
\eea
For $\beta=1$, we may consider all variables $S_{i,m}$ to be independent variables, and we integrate by parts:
\bea
&& \sum_{l\neq j}  {M_{i,l}-M_{i,j}\over y_l-y_j}  \cr
&=& \sum_{l\neq j}\sum_m\,{\beta\over x_i}\,\det(X)^{1-\beta}\, \int dS\,\, {\ee{\Tr SX}\,\over \prod_{k=1}^n \det(S-y_k )^{\beta}} \,\,((S-y_l )^{-1})_{i,m}((S-y_j )^{-1})_{m,i} \cr
&=& -\sum_m\,{1\over x_i}\,\det(X)^{1-\beta}\, \int dS\,{\ee{\Tr SX}\over \det(S-y_j)^\beta}\,((S-y_j )^{-1})_{m,i}\,{\d\over \d S_{i,m}}\, {1\over \prod_{l\neq j} \det(S-y_l )^{\beta}}  \cr
&=&  \sum_m\,{1\over x_i}\,\det(X)^{1-\beta}\, \int dS\,\,
{1\over \prod_{l\neq j} \det(S-y_l )^{\beta}}\, {\d\over \d S_{i,m}}\,    {\ee{\Tr SX}\over \det(S-y_j)^\beta}\,((S-y_j )^{-1})_{m,i}  \cr
&=&  \sum_m {1\over x_i} \det(X)^{1-\beta}\, \int dS\,\,
{\ee{\Tr SX}\over \prod_{k} \det(S-y_k )^{\beta}} \,  ((S-y_j )^{-1})_{m,i} \, x_i \delta_{i,m}  \cr
&& -\beta \sum_m\,{1\over x_i}\,\det(X)^{1-\beta}\, \int dS\,\,
{\ee{\Tr SX}\over \prod_{k} \det(S-y_k )^{\beta}} \,  ((S-y_j )^{-1})_{i,m}\,((S-y_j )^{-1})_{m,i}  \cr
&& - \sum_m\,{1\over x_i}\,\det(X)^{1-\beta}\, \int dS\,\,
{\ee{\Tr SX}\over \prod_{k} \det(S-y_k )^{\beta}} \, ((S-y_j )^{-1})_{m,m} ((S-y_j )^{-1})_{i,i}  \cr
&=&   \det(X)^{1-\beta}\, \int dS\,\,
{\ee{\Tr SX}\over \prod_{k} \det(S-y_k )^{\beta}} \,  ((S-y_j )^{-1})_{i,i}    \cr
&& -\,{1\over x_i}\,\det(X)^{1-\beta}\, \int dS\,\,
{\ee{\Tr SX}\over \prod_{k} \det(S-y_k )^{\beta}} \,  ((S-y_j )^{-2})_{i,i}  \cr
&& - \,{\beta\over x_i}\,\det(X)^{1-\beta}\, \int dS\,\,
{\ee{\Tr SX}\over \prod_{k} \det(S-y_k )^{\beta}} \, \Tr {(S-y_j )^{-1}} ((S-y_j )^{-1})_{i,i}  \cr
&=& {1\over \beta}\,\,\left(x_i M_{i,j} - {\d M_{i,j}\over \d y_j} \right)
\eea
QED.
The same computation can be repeated for $\beta=1/2$ and $\beta=2$, with additional steps because the variables $S_{i,m}$ are no longer independent, and also because for $\beta=2$, $\det (S-y_j)$ is defined as the product of singular values.
Another proof is given in section \ref{secproofDunklloop}.

\medskip

{\bf Remark:}
Of course a similar equation can be found by exchanging the roles of $X$ and $Y$, and one gets the symmetric linear equation:
\beq\label{eqdiflinMijXtoY}
\forall i,j \virg
{\partial M_{i,j}\over \partial x_i} + \beta \sum_{l\neq i} {M_{l,j}-M_{i,j}\over x_l-x_i}  = M_{i,j}\, y_j
\eeq

\medskip

{\bf Remark:}
again, this proves that \eq{MijintdS} is solution of the differential equation \eq{eqdiflinMij}, for any choice of integration domain provided that we can integrate by parts.
In fact, by taking linear combinations of all possible integration contours, we get the general solution of the linear equation \eq{eqdiflinMij}.
However, a general solution of \eq{eqdiflinMij} is not necessarily symmetric in $X$ and $Y$, and does not necessarily obey \eq{eqdiflinMijXtoY}.

\subsection{Calogero equation}

Here, we prove that $I_{\beta,n}$ satisfies the Calogero equation.

Start from the linear equation:
\beq
{\partial M_{i,j}\over \partial x_i} + \beta \sum_{k\neq i} {M_{i,j}-M_{k,j}\over x_i-x_k}  = M_{i,j}\, y_j
\eeq
then sum over $j$, using \eq{IsumMij}:
\beq
{\partial I \over \partial x_i} = \sum_j M_{i,j}\, y_j
\eeq
Then apply ${\partial  \over \partial x_i}$:
\bea
{\partial^2 I \over \partial x_i^2} 
&=& \sum_j y_j\,\, {\partial M_{i,j}\over \partial x_i} \cr
&=& \sum_j y_j\,\, \Big(  M_{i,j}\, y_j - \beta \sum_{k\neq i} {M_{i,j}-M_{k,j}\over x_i-x_k}  \Big) \cr
&=& \sum_j y_j^2 \,\,  M_{i,j} - \beta  \sum_j y_j\,\, \sum_{k\neq i} {M_{i,j}-M_{k,j}\over x_i-x_k}   \cr
&=& \sum_j y_j^2 \,\,  M_{i,j} - \beta  \sum_{k\neq i} {{\partial I\over \partial x_i}-{\partial I \over \partial x_k}\over x_i-x_k}   \cr
\eea
If we take the sum over $i$, using \eq{IsumMijXtoY}, we get:
\beq
\sum_i {\partial^2 I \over \partial x_i^2} 
+ \beta  \sum_i \sum_{k\neq i} {{\partial I\over \partial x_i}-{\partial I \over \partial x_k}\over x_i-x_k}   
= \sum_j y_j^2 \,\,  I 
\eeq
i.e. we recover the Calogero equation:
\beq
\encadremath{
H_{\rm Calogero}.I_{\beta,n} = (\sum_j y_j^2)\,I_{\beta,n}
}\eeq

\subsection{Matrix form of the linear equations}

The linear equations, are $n^2$ linear equations of order 1, for $n^2$ unknown functions $M_{i,j}$.
They can be summarized into a matricial equation:
\beq
M\, Y = K\, M
\eeq
where $K$ is a matricial operator
\beq
K_{ii} =\frac{\partial }{\partial x_{i}}+\beta \sum_{k\neq i}\frac{1}{\left( x_{i}-x_{k}\right) } 
\virg
K_{ik} =-\frac{\beta }{\left( x_{i}-x_{k}\right) }\ \ \ \ \ \ \ \ i\neq k
\eeq
and more generally this implies:
\beq
M\, Y^p = K^p\, M
\eeq
and therefore, for any polynomial $P$:
\beq
M. P(Y)= P(K). M
\eeq
In particular if we choose the characteristic polynomial of $Y$:
\beq
0 = \prod_{i=1}^n (y_i-K)\,\, . \, M 
\eeq

\bigskip
Let us introduce the vector
\beq
e=(1,1,\dots,1)^t
\eeq
It is such that $M$ is a stochastic matrix, i.e.:
\beq
M.e=I_{\beta,n}\,\, e
\virg
e^t . M =  I_{\beta,n}\,\, e^t
\eeq
We thus have, for any polynomial $P$:
\beq
e^t\,\, P(K)\,\, e\, . \, I_{\beta,n} = I_{\beta,n}\,\,\, \Tr P(Y)
\eeq
Notice that the Calogero equation is the case $P(K)=K^2$.

If $P$ is the characteristic polynomial of $Y$ we get another differential equation for $I_{\beta,n}$:
\beq
\encadremath{
\forall \, i \, , \qquad \quad \sum_j\, \left(\prod_{l=1}^n (y_l-K)\right)_{i,j}\,\, . \, I_{\beta,n} = 0
}\eeq
And if $P(K) = \prod_{l\neq j} (y_l-K)$, we get:
\beq
\encadremath{
M_{i,j} = \sum_m \left(\prod_{l\neq j} {y_l-K\over y_l-y_j}\right)_{i,m} \, .\, I_{\beta,n} 
}\eeq
This last relation allows to reconstruct $M_{i,j}$ if we know $I_{\beta,n}$.

\medskip

Finally, before leaving this section, we just mention that those operators $K_{i,j}$ are also related to the Laplacian over the set of matrices $E_{\beta,n}$, as was noted recently by Zuber \cite{Zuber}.

\subsection{Linear equation from loop equations}
\label{secproofDunklloop}

There is another way of deriving those Dunkl-like linear equations for the angular integrals, using loop equations of an associated 2-matrix model.

Consider the following 2-matrix integral, where $M_1$ and $M_2$ are both in the $E_{\beta,n}$ ensemble:
\beq
Z = \int dM_1 \, dM_2\,\, \ee{-\Tr ( V_1(M_1)+V_2(M_2)-M_1 M_2)}
\eeq
After diagonalization of $M_1 = O_1 X O_1^{-1}$ and $M_2=O_2 Y O_2^{-1}$, we have:
\beq
Z = \int dX dY dO_1 dO_2\,\, \Delta(X)^{2\beta}\Delta(Y)^{2\beta}\,\, \ee{-\Tr ( V_1(X)+V_2(Y))}\,\,\, \ee{\Tr X O_1^{-1}O_2 Y O_2^{-1} O_1}
\eeq
We redefine $O_2=O_1. O$, and the integral over $O_1$ gives $1$, and the integral over $O$ gives the angular integral:
\beq
Z = \int dX dY \,\, \Delta(X)^{2\beta}\Delta(Y)^{2\beta}\,\, \ee{-\Tr ( V_1(X)+V_2(Y))}\,\,\, I_{\beta,n}(X,Y)
\eeq
We can do a similar change of variable for moments:
\bea
&& < \Tr {1\over x-M_1}\, {1\over y-M_2}> \cr
&=& {1\over Z}\,\, \int dM_1 \, dM_2\,\, \ee{-\Tr ( V_1(M_1)+V_2(M_2)-M_1 M_2)}\,\, \Tr \left( {1\over x-M_1}\, \, {1\over y-M_2} \right) \cr
&=& {1\over Z}\,\, \int dX dY dO_1 dO_2\,\, \Delta(X)^{2\beta}\Delta(Y)^{2\beta}\,\, \ee{-\Tr ( V_1(X)+V_2(Y))}\,\,\, \cr
&& \ee{\Tr X O_1^{-1}O_2 Y O_2^{-1} O_1}\,\, \Tr \left( {1\over x-X}\, O_1^{-1}O_2 \, {1\over y-Y} O_2^{-1} O_1 \right) \cr
&=& {1\over Z}\,\, \int dX dY \,\, \Delta(X)^{2\beta}\Delta(Y)^{2\beta}\,\, \ee{-\Tr ( V_1(X)+V_2(Y))}\,\,\, \cr
&& \qquad \quad \sum_{i,j} M_{i,j}(X,Y)\,\, {1\over x-x_i}\,\, \, {1\over y-y_j}  \cr
\eea
where $M_{i,j}(X,Y)$ is the Morozov moment defined in eq.\ref{defMorozovmoments}.

Loop equations amount to say that an integral is invariant under a change of variables.
Thus, we change $M_1\to M_1+\epsilon {1\over x-M_1}\,{1\over y-M_2}+O(\epsilon^2)$ in $Z$, and to order $1$ in $\epsilon$ we get (the loop equations for $\beta=1/2,1,2$ ensembles can be found in several references \cite{}, the Jacobian is easily computed in eigenvalue representation, see appendix \ref{apploopeq}, \eq{loopeqbeta2app}):
\bea
0
&=& \left<\Tr {1\over x-M_1}\, {M_2\over y-M_2}\right> 
-\left<\Tr {V_1'(M_1)\over x-M_1}\, {1\over y-M_2}\right>  \cr
&& +\beta \left<\Tr {1\over x-M_1} \Tr {1\over x-M_1}\,{1\over y-M_2}\right> \cr
&& +(\beta-1){\partial \over \partial x} \left<\Tr  {1\over x-M_1}\,{1\over y-M_2} \right>  \cr
\eea
i.e., going to eigenvalues $M_1 = O_1 X O_1^{-1}$ and $M_2=O_2 Y O_2^{-1}$:
\bea
0
&=& \sum_{i,j} \left< {(y_j-V'_1(x_i))\,M_{i,j}(X,Y)\over (x-x_i)(y-y_j)}\right>  \cr
&& +\beta \sum_{i\neq l}\sum_j \left< {M_{i,j}(X,Y)\over (x-x_l)(x-x_i)(y-y_j)} \right> \cr
&& + \sum_{i,j}  \left<{M_{i,j}(X,Y)\over (x-x_i)^2(y-y_j)} \right>  \cr
\eea
The last term can be integrated by parts:
\bea
&& \sum_{i,j}  \left<{M_{i,j}(X,Y)\over (x-x_i)^2(y-y_j)} \right> \cr
&= & \sum_{i,j} \int dX\, dY\,\, \Delta(X)^{2\beta}\Delta(Y)^{2\beta}\,\, \ee{-\Tr ( V_1(X)+V_2(Y))}\,\,\cr
&& \qquad M_{i,j}(X,Y)\,\,\, {\partial \over \partial x_i}\,\, {1\over (x-x_i)\,(y-y_j)}  \cr
&= & - \sum_{i,j} \int dX\, dY\,\, {1\over (x-x_i)\,(y-y_j)}\,\,\, \cr
&& \qquad {\partial \over \partial x_i}\,\,M_{i,j}(X,Y)\,\, \Delta(X)^{2\beta}\Delta(Y)^{2\beta}\,\, \ee{-\Tr ( V_1(X)+V_2(Y))}  \cr
&= &  \sum_{i,j} \int dX\, dY\,\, {M_{i,j}(X,Y)\over (x-x_i)\,(y-y_j)}\,\,\,  \Delta(X)^{2\beta}\Delta(Y)^{2\beta}\,\, \cr
&& \qquad \ee{-\Tr ( V_1(X)+V_2(Y))}  
\,\, \Big( V'_1(x_i) - \sum_{l\neq i} {2\beta\over x_i-x_l}  \Big) \cr
&& - \sum_{i,j} \int dX\, dY\,\, {1\over (x-x_i)\,(y-y_j)}\,\,\,  \Delta(X)^{2\beta}\Delta(Y)^{2\beta}\,\, \cr
&& \qquad \ee{-\Tr ( V_1(X)+V_2(Y))}  \,\, {\partial M_{i,j}(X,Y)\over \partial x_i} \cr
&= & \sum_{i,j}  \left<{V'_1(x_i)M_{i,j}(X,Y)\over (x-x_i)(y-y_j)} \right> 
-2\beta  \sum_{l\neq i}\sum_j  \left<{M_{i,j}(X,Y)\over (x_i-x_l)(x-x_i)(y-y_j)} \right> \cr
&& - \sum_{i,j} \left< {1\over (x-x_i)(y-y_j)}\,\,{\partial M_{i,j}(X,Y)\over \partial x_i}\right> \cr
\eea
Therefore we have:
\bea
&&  \sum_{i,j} \left< {1\over (x-x_i)(y-y_j)}\,\,{\partial M_{i,j}(X,Y)\over \partial x_i}\right> \cr
&= & \sum_{i,j}  \left<{y_j M_{i,j}(X,Y)\over (x-x_i)(y-y_j)} \right> 
-2\beta  \sum_{l\neq i}\sum_j  \left<{M_{i,j}(X,Y)\over (x_i-x_l)(x-x_i)(y-y_j)} \right> \cr
&& +\beta \sum_{i\neq l}\sum_j \left< {M_{i,j}(X,Y)\over (x-x_l)(x-x_i)(y-y_j)} \right> \cr
&= & \sum_{i,j}  \left<{y_j M_{i,j}(X,Y)\over (x-x_i)(y-y_j)} \right> 
-2\beta  \sum_{l\neq i}\sum_j  \left<{M_{i,j}(X,Y)\over (x_i-x_l)(x-x_i)(y-y_j)} \right> \cr
&& +\beta \sum_{i\neq l}\sum_j \left< {M_{i,j}(X,Y)\over (x-x_l)(x_l-x_i)(y-y_j)} \right>  \cr
&& +\beta \sum_{i\neq l}\sum_j \left< {M_{i,j}(X,Y)\over (x-x_i)(x_i-x_l)(y-y_j)} \right> \cr
&= & \sum_{i,j}  \left<{y_j M_{i,j}(X,Y)\over (x-x_i)(y-y_j)} \right> \cr
&& -\beta  \sum_{l\neq i}\sum_j  \left<{1\over (x-x_i)(y-y_j)}\,\,{M_{i,j}(X,Y)-M_{l,j}(X,Y)\over (x_i-x_l)} \right> \cr
\eea
Since this equation must hold for any $V_1$ and $V_2$, $x$, $y$, i.e. for any measure on $X$ and $Y$, it must hold term by term i.e. we recover the linear equation:
\beq
  {\partial M_{i,j}(X,Y)\over \partial x_i} 
= y_j\, M_{i,j}(X,Y)
-\beta  \sum_{l\neq i}\,\,{M_{i,j}(X,Y)-M_{l,j}(X,Y)\over (x_i-x_l)}  
\eeq
Of course, the loop equation coming from the change of variable $M_2\to M_2+\epsilon {1\over x-M_1}\,{1\over y-M_2}+O(\epsilon^2)$ gives the symmetric linear equation:
\beq
  {\partial M_{i,j}(X,Y)\over \partial y_j} 
= x_i\, M_{i,j}(X,Y)
-\beta  \sum_{l\neq j}\,\,{M_{i,j}(X,Y)-M_{i,l}(X,Y)\over (y_j-y_l)}  
\eeq
QED.

\section{Principal terms and the $\tau_{ij}$ variables}
\label{secpterms}

As we mentioned in the introduction, it was noticed in particular by Brezin and Hikami \cite{BrezHikbeta}, that the angular integral can be written as combinations of exponential terms, and polynomials (for $\beta$ integer, series otherwise), of some reduced variables $\tau_{i,j}=-{1\over2} \,(x_i-x_j)(y_i-y_j)$.
Here, we show how our recursion gives such a form.

We thus define:

\bd

We define the principal term $\Ibar_{\beta,n}(X;Y)$ from the recursion:
\bea\label{recIbardef}
 \Ibar_{\beta,1} &=& 1  \cr
& {\rm and}& \cr
 \Ibar_{\beta,n}(X_n;Y_n) 
&=&  \Delta(Y_n)^{2\beta}\prod_{i=1}^{n-1} (x_i-x_n)
 \,\,\Res_{\l_i\to y_i} {d\l_1\over (\l_1-y_1)^\beta}\,\dots {d\l_{n-1}\over (\l_{n-1}-y_{n-1})^\beta} \cr
 && \quad \,\,\,\, {\Ibar_{\beta,n-1}(X_{n-1},\L)\,\,\ee{\sum_i (x_i-x_n)(\l_i-y_i)}\over \prod_{k=1}^n\prod_{i=1,\neq k}^{n-1} (y_k-\l_i)^{\beta}} \cr
\eea
It is such that (the sum over permutations comes from the sum of residues at all poles in recursion \eq{mainrecIbetan} of theorem. \ref{thmainrecIbetan}):
\beq
\encadremath{
I_{\beta,n}(X,Y) = \sum_{\sigma} {\ee{\sum_{i=1}^n x_i y_{\sigma(i)}}\over \Delta(X)^{2\beta}\Delta(Y_\sigma)^{2\beta}}\,\,\, \Ibar_{\beta,n}(X,Y_\sigma)
}\eeq
\ed

In \cite{BrezHikbeta}, Brezin and Hikami observed and conjectured that when $\beta$ is an integer, $\Ibar_{\beta,n}(X,Y)$ is a polynomial in the variables $\tau_{i,j}=-{1\over 2}(x_i-x_j)(y_i-y_j)$.

\medskip

For instance, if $\beta=1$, we have for arbitrary $n$:
\beq
\Ibar_{1,n} = \prod_{i<j} \tau_{i,j}
\eeq
And, if $n=2$, we have for arbitrary $\beta$:
\bea
\Ibar_{\beta,2} 
&=& (x_1-x_2)\,(y_1-y_2)^{2\beta}\,\, \Res_{\l\to 0} {d\l\over \l^\beta}\,\,{\ee{(x_1-x_2)\l}\over (y_2-y_1-\l)^\beta} \cr
&=& (x_1-x_2)\,(y_1-y_2)^{2\beta}\,\, {\partial^{\beta-1}\over \partial\l^{\beta-1}}\,\,\, \left({\ee{(x_1-x_2)\l}\over (y_2-y_1-\l)^\beta}\right)_{\l=0} \cr
&=& (x_1-x_2)\,(y_1-y_2)^{2\beta}\,\, \sum_{k=0}^{\beta-1} \cr
&& \qquad {(\beta-1)!\over k! (\beta-1-k)!}\,\, (x_1-x_2)^{\beta-1-k}\,\, (y_2-y_1)^{-\beta-k}\,\, {(\beta-1+k)!\over (\beta-1)!} \cr
&=&  \sum_{k=0}^{\beta-1} {(\beta-1+k)!\over k! (\beta-1-k)!}\,\, \left((x_1-x_2)(y_2-y_1)\right)^{\beta-k} \cr
&=& 2^{\beta}\,\, Q_{\beta,0}(\tau_{1,2})
\eea
i.e. we recover the well known result that  $\Ibar_{\beta,2}$ is the Bessel polynomial of degree $\beta$.

We are going to prove the conjecture of Brezin Hikami for all $n$ and for all $\beta\in{\mathbb N}$, but first, let us prove some preliminary properties:

\bl\label{lemmaIbarpolsym}
$\Ibar_{\beta,n}$ is a polynomial in all variables $x_i$ and $y_j$, and it is symmetric under the exchange $X \leftrightarrow Y$, and under the permutation of pairs $(x_i,y_i)\leftrightarrow (x_j,y_j)$, and under translations $X\to X+{\rm cte}.{\rm Id}$ or $Y\to Y+{\rm cte}.{\rm Id}$.

\el
\proof{
If $\beta$ is an integer, the recursion relation \eq{recIbardef} leads to (we write $x_{i,j}=x_i-x_j$, $y_{i,j}=y_i-y_j$):
\bea
 \Ibar_{\beta,n+1} 
&=&   \prod_{i=1}^n x_{i,n+1}\, y_{n+1,i}^\beta\, \cr
&& \left(\partial \over \partial \l_i\right)^{\beta-1}\, \Big[\ee{x_{i,n+1}\l_i}\,\prod_{1\leq j\leq n+1,j \neq i}\, {y_{j,i}^\beta\over (y_{j,i}-\l_i)^\beta} \Big]\,\,\,\, \Ibar_{\beta,n}(X_n,Y_n+\L)\,\, \Big|_{\L=0} \cr
 \eea
which shows by recursion, that $\Ibar_{\beta,n+1}$ is a rational function of all $x_i$'s and $y_j$'s.

We know, from its very definition, that the angular integral
\beq
I_{\beta,n}(X,Y) = \sum_{\sigma} {\ee{\sum_{i=1}^n x_i y_{\sigma(i)}}\over \Delta(X)^{2\beta}\Delta(Y_\sigma)^{2\beta}}\,\,\, \Ibar_{\beta,n}(X,Y_\sigma)
\eeq
is symmetric in all $x_i$'s and $y_j$'s, and in the exchange $X\leftrightarrow Y$.
Since the exponentials are linearly independent on the ring of rational functions, each term must be symmetric in permutations of pairs $(x_i,y_i)$'s, i.e. $\Ibar_{\beta,n}(x_1,\dots,x_n;y_1,\dots,y_n)$ is a symmetric function of the pairs $(x_i,y_i)$'s, and also symmetric under $X \leftrightarrow Y$.

Moreover, $\Ibar_{\beta,n+1}$ is clearly a polynomial in the variables $x_{n+1}$ and $y_{n+1}$, and because of the symmetry, it must also be a polynomial in all variables.
Translation invariance is also clear from the recursion formula.

}

\bt\label{thBH} (Conjecture of Brezin-Hikami):

$\Ibar_{\beta,n}$ is a symmetric polynomial of degree $\beta$ in the $\tau_{i,j}$'s.
\et

\medskip

\proof{
Using lemma \ref{lemmaIbarpolsym}, it is easy to see that $\Ibar_{\beta,n}$ fulfills the hypothesis of lemma \ref{LemmaPsymtau} in the appendix \ref{applemmaPoltau}, and this proves the theorem.
}
\bigskip

I.e. we have proved the conjecture of Brezin and Hikami \cite{BrezHikbeta}.
In fact, we notice that the property of being a polynomial in the $\tau$'s, is not specific to angular integrals, but comes only from the global symmetries.

\subsection{Recursion without residues for $\beta$ integer}

For $\beta\in {\mathbb N}$, the residues in recursion \eq{recIbardef} can be performed, and they compute derivatives of the integrand. Thus \eq{recIbardef} can be rewritten:
\bea
 && \Ibar_{\beta,n}(X_n;Y_n) \cr
&=&  \Delta(Y_n)^{2\beta}\prod_{i=1}^{n-1} (x_i-x_n)
 \,\,\Res_{\l_i\to a_i} {d\l_1\over (\l_1-a_1)^\beta}\,\dots {d\l_{n-1}\over (\l_{n-1}-a_{n-1})^\beta} \cr
 && \quad \,\,\,\, {\Ibar_{\beta,n-1}(X_{n-1},\L)\,\,\ee{\sum_i (x_i-x_n)(\l_i-y_i)}\over \prod_{k=1}^n\prod_{i=1,\neq k}^{n-1} (y_k-\l_i)^{\beta}} \Big|_{a_i=y_i} \cr
&=& {\Delta(Y_n)^{2\beta}\over (\beta-1)!^{n-1}}\,\, \prod_{i=1}^{n-1} (x_i-x_n)\,\,\,
 \prod_i \left(\partial\over\partial a_i\right)^{\beta-1} \,\,\Res_{\l_i\to a_i} {d\l_1\over (\l_1-a_1)}\,\dots {d\l_{n-1}\over (\l_{n-1}-a_{n-1})} \cr
 && \quad \,\,\,\, {\Ibar_{\beta,n-1}(X_{n-1},\L)\,\,\ee{\sum_i (x_i-x_n)(\l_i-y_i)}\over \prod_{k=1}^n\prod_{i=1,\neq k}^{n-1} (y_k-\l_i)^{\beta}} \Big|_{a_i=y_i} \cr
\eea
i.e. we can perform the residues:
\beq\label{recsansres}
\encadremath{
 \Ibar_{\beta,n}(X_n;Y_n) 
= {\Delta(Y_n)^{2\beta}\over (\beta-1)!^{n-1}}\,\, \prod_{i=1}^{n-1} x_{i,n}\,\,\,
  \left(\partial\over\partial a_i\right)^{\beta-1} \,\, {\Ibar_{\beta,n-1}(X_{n-1},a)\,\,\ee{\sum_i x_{i,n}(a_i-y_i)}\over \prod_{k=1}^n\prod_{i=1,\neq k}^{n-1} (y_k-a_i)^{\beta}} \Big|_{a_i=y_i} 
}\eeq
More explicitly
\bea
&& \Ibar_{\beta,n}(X_n;Y_n) \cr
&=& 
\prod_{i=1}^{n-1}
\sum_{\gamma_i, \beta_{i,k}=0}^{\beta-1} \,\, {(x_{i,n}y_{n,i})^{\beta_{i,n}}\over  \Gamma(\beta-\gamma_i-\sum_k \beta_{i,k})\,\, \beta_{i,n}!}\,\, \prod_{k=1,k\neq i}^{n-1} {\Gamma(\beta+\beta_{i,k})\over \beta_{i,k}!\,\Gamma(\beta)}\,\, {1\over (x_{i,n}y_{k,i})^{\beta_{i,k}}} \cr
&& {1\over \gamma_i!}\,\left({1\over x_{i,n}}\,{\partial\over \partial a_i}\right)^{\gamma_i}
\qquad \Ibar_{\beta,n-1}(X_{n-1},a)\,\, \Big|_{a_i=y_i} \cr
\eea

%


%

%


%



\subsection{$n=3$}
\label{secn3}

For $n=3$, and arbitrary $\beta$, we prove that:
\bt
\beq
\encadremath{
\Ibar_{\beta,3} =   \,\sum_{k=0}^{\infty} {\Gamma(\beta-k)\over 2^{3k}\, k!\,\Gamma(\beta+k)}\, \prod_{i<j} Q_{\beta,k}(\tau_{ij})  
}
\eeq
\et
In fact, for $\beta$ integer, the sum over $k$ is finite and reduces to $k\leq \beta-1$.

\medskip
The proof, rather technical,  is given in appendix \ref{appproofn3}. We used the Calogero equation.

\subsection{Conjecture for higher $n$}

Applying the recursion relations of this article, we also computed the $n=4$ case for small values of $\beta$:
\bea
\Ibar_{2,4} 
&= &   \prod_{i<j} Q_{2,0}(\tau_{ij})  \cr
& & +   {1\over 16}\, Q_{2,0}(\tau_{1,2})Q_{2,0}(\tau_{1,3}) Q_{2,0}(\tau_{1,4}) \,\,Q_{2,1}(\tau_{2,3}) Q_{2,1}(\tau_{2,4})Q_{2,1}(\tau_{3,4}) + {\rm sym}\cr
& & +  {1\over 128}\, Q_{2,0}(\tau_{1,2}) \,\, Q_{2,1}(\tau_{1,3}) Q_{2,1}(\tau_{1,4}) Q_{2,1}(\tau_{2,3}) Q_{2,1}(\tau_{2,4})Q_{2,1}(\tau_{3,4}) + {\rm sym} \cr
\eea
and:
\bea
\Ibar_{3,4} 
&= &   64 \prod_{i<j} Q_{3,0}(\tau_{ij})  \cr
& & +   {4\over 3}\, Q_{3,0}(\tau_{1,2})Q_{3,0}(\tau_{1,3}) Q_{3,0}(\tau_{1,4}) \,\,Q_{3,1}(\tau_{2,3}) Q_{3,1}(\tau_{2,4})Q_{3,1}(\tau_{3,4}) + {\rm sym}\cr
& & +   {1\over 48}\, Q_{3,0}(\tau_{1,2})Q_{3,0}(\tau_{1,3}) Q_{3,0}(\tau_{1,4}) \,\,Q_{3,2}(\tau_{2,3}) Q_{3,2}(\tau_{2,4})Q_{3,2}(\tau_{3,4}) + {\rm sym}\cr
&& + \dots
\eea
Those expressions lead us to conjecture a general form in terms of Bessel polynomials:

\begin{conjecture}\label{conjhatIbetan}

We conjecture that for all $n$ and $\beta$, $\hat{I}_{\beta,n}$ is of the form:
\beq
\encadremath{
\Ibar_{\beta,n} = \sum_{\{l\}} A_{\{ l\}}\,\, \prod_{i<j}\, Q_{\beta,l_{(i,j)}}(\tau_{i,j})
}\eeq

\end{conjecture}
Unfortunately, we have not been able so far to determine the general form of the coefficients $A_{\{ l\}}$  for $n>3$ (except $n=4$ and $\beta=2$).

\subsection{Symplectic case $\beta=2$}
\label{secbeta2}

For $\beta=2$, the recursion \eq{recsansres}, reduces to:
\bea\label{recI2ndda}
\Ibar_{2,n} 
&=& {\Delta(Y_n)^{4}}\,\, \prod_{i=1}^{n-1} x_{i,n}\,\,\,
\prod_i {\partial\over\partial a_i} \,\, {\Ibar_{2,n-1}(X_{n-1},a)\,\,\ee{\sum_i (x_i-x_n)(a_i-y_i)}\over \prod_{k=1}^n\prod_{i=1,\neq k}^{n-1} (y_k-a_i)^2} \Big|_{a_i=y_i} \cr
&=&  \prod_{i=1}^{n-1} x_{i,n}\, y_{i,n}^2\,\,
 \Big(
x_i-x_n - \sum_{k=1, \neq i}^{n} {2\over y_i-y_k} + {\partial\over\partial a_i}
\Big)
\Ibar_{2,n-1}(X_{n-1},a)\,\, \Big|_{a_i=y_i} \cr
\eea
From a recursion hypothesis, we assume that $\Ibar_{2,n-1}(X_{n-1},a)$ is a polynomial in the $\tau$'s of the form:
\beq\label{I2tauQhyp}
\Ibar_{2,n-1}(X_{n-1},Y_{n-1}) = \sum_{\{ l\}} A_{\{l\}}\,\, \prod_{i<j} Q_{2,l_{i,j}}(\tau_{i,j})
\eeq
where for every pair $(i,j)$ we have $l_{i,j}\in\{0,1\}$. We recall that:
\beq
|0>_\tau=Q_{2,0}(\tau) = \tau^2+\tau
\virg
|1>_\tau=Q_{1,0}(\tau) = 2\tau
\eeq

Thus we may write:
\beq
{\partial\over\partial a_i} = -\,{1\over 2} \,\sum_{k\neq i} x_{i,k} {\partial\over\partial t_{i,k}}
\virg
t_{i,k} = -{1\over 2}\,(x_i-x_k)(a_i-a_k)
\eeq
and we define the operators $C_{i,k}$ acting on functions of the variable $\tau_{i,k}$ such that:
\beq
C_{i,k} = {1\over \tau_{i,k}} -{1\over 2}  {\partial \over \partial t_{i,k}}
\eeq
and all derivatives must be eventually computed at $t_{i,k}=\tau_{i,k}$.

Since our operators act on expressions of the form \eq{I2tauQhyp}, we need to compute:
\beq
C_{i,j}|0> = C_{i,j}.Q_{2,0}(\tau_{i,j}) = C_{i,j}.(\tau_{i,j}^2+\tau_{i,j}) = {1\over 2}
\eeq
\beq
C_{i,j}|1> = C_{i,j}.Q_{2,1}(\tau_{i,j}) = C_{i,j}.\, 2\tau_{i,j} = 1
\eeq
And we may also have terms of the form $C_{i,j}.C_{j,i}$, for which we have:
\beq
C_{i,j} C_{j,i} \, |0> = C_{i,j} C_{j,i} \, .(\tau_{i,j}^2+\tau_{i,j}) = - {1\over 2}
\virg
C_{i,j} C_{j,i} \, |1> = C_{i,j} C_{j,i} \, .\tau_{i,j} = 0
\eeq

Finally we have:
\beq
\Ibar_{2,n}
=  \prod_{i=1}^{n-1} \tau_{i,n}^2\,\,\,\prod_{i=1}^{n-1}
 \Big(
1+{1\over \tau_{i,n}} +  \sum_{k=1, \neq i}^{n-1} {x_{i,k} \over x_{i,n}}\,C_{i,k}\Big)
\qquad \Ibar_{2,n-1}  \Big|_{t_{i,j}=\tau_{i,j}}
\eeq
It is more convenient to rewrite this in terms of a Hilbert space with basis $|0>=Q_{2,0}$ and $|1>=Q_{2,1}$, and thus:
\bea
\Ibar_{2,n}
&=&  \prod_{i=1}^{n-1} 
 \Big(
\tau_{i,n}^2+\tau_{i,n} + {1\over 2}\,\tau_{i,n} \sum_{k=1, \neq i}^{n-1} {y_{n,i}\over y_{k,i}}\, A_{i,k}\Big)
\qquad \Ibar_{2,n-1}  \cr
\eea
where
\beq
A_{i,k} = 2\tau_{i,k}C_{i,k}
\eeq
We have:
\beq
A |0> = {1\over 2} |1>
\virg
A |1> =  |1>
\eeq
and:
\beq
A_{i,k}A_{k,i} = 2(A_{i,k}-1)
\eeq

Unfortunately we have not been able to go further with this formulation.

\subsubsection{Triangle conjecture}

We have seen that there is an operator formalism for computing angular integrals, with operators $A_{i,j}$ associated to "edges" $(i,j)$.
However, it can be seen for $n=3,4$, that operator edges appear only in certain combinations, which involve triangles $(i,j,k)$.
We thus introduce the triangles operators:
\beq
T_{i,j,k} = \pi_{i,j}\pi_{j,k}\pi_{k,i}
\eeq
where $\pi_{i,j}=\pi_{j,i}$ is the projector on state ${1\over 2}|1>_{i,j}=\tau_{i,j}$:
\beq
\pi_{i,j}.(\tau_{i,j}^2+\tau_{i,j}) = \tau_{i,j}
\virg
\pi_{i,j}.\tau_{i,j}=\tau_{i,j}
\virg \pi_{i,j}=\pi_{j,i}
\virg \pi_{i,j}^2=\pi_{i,j}
\eeq

With this notation we have:
\beq
\Ibar_{2,3} = (1+{1\over 2}T_{1,2,3}).\prod_{1\leq i<j\leq 3} |0>_{i,j}
\eeq
where we recall that $|0>_{i,j} = \tau_{i,j}^2+\tau_{i,j}$.

And
\bea
\Ibar_{2,4} 
&=& \Big(1+{1\over 2}\, (T_{1,2,3}+T_{1,2,4}+T_{1,3,4}+T_{2,3,4}) \cr
&& + {1\over 4}(T_{1,2,3}T_{1,2,4}+T_{1,2,3}T_{1,3,4}+T_{1,2,4}T_{1,3,4}\cr
&& +T_{1,2,3}T_{2,3,4}+T_{1,2,4}T_{2,3,4}+T_{1,3,4}T_{2,3,4}) \Big). \prod_{1\leq i<j\leq 4} |0>_{i,j}
\eea

We are naturally led to conjecture that:
\beq
\Ibar_{2,n}
= \Big(\sum_{{\rm triangulations}\, {\cal T}} C_{{\cal T}}\, \prod_{T\in {\cal T}}\Big). \prod_{i<j} |0>_{i,j}
\eeq
We have:
\beq
C_{\emptyset}=1
\virg
C_{(i,j,k)}={1\over 2}
\virg
C_{(i,j,k),(i,j,l)}={1\over 4}
\virg \dots
\eeq

We have not been able so far to prove this conjecture.
It is to be noted from the low values of $n$, that triangles seem to play a role for all $\beta\in{\mathbb N}$.

\subsubsection{Additional results: determinantal recursion}

Just for completeness, we give another form of the recursion \eq{recI2ndda}, in terms of determinants:
%
%
\beq\label{rechatIdet}
 \hat{I}_{2,n+1}(X_{n+1};Y_{n+1}) 
 =  {\det\left[ X-x_{n+1}-{2\over Y-y_{n+1}}+B+\partial_Y \right]\over \det(X-x_{n+1})}
 \hat{I}_{2,n}(X_{n};Y_{n}) 
\eeq
where
\beq
\hat{I}_{2,n}(X,Y) = {1\over \prod_{i<j} \tau_{i,j}^2}\,\, \Ibar_{2,n}(X,Y)
\eeq
and where $B$ is the antisymmetric matrix
\beq
B_{ij}={\sqrt{2}\over y_i-y_j}
\eeq
and $\partial_Y={\rm diag}(\partial_{y_1},\dots,\partial_{y_n})$.
This is proved by observing that the expansion of the determinant \eq{rechatIdet}
can be interpreted like a Wick's expansion equivalent to \eq{recI2ndda}.

\section{Conclusion}
\label{secconcl}

In this article, we have found many new relations and new representations of angular integrals.

First, we have been able to rewrite angular integrals with a complicated Haar measure, in terms of usual Lebesgue measure contour integrals.
Then, we have deduced duality and recursion formulae.

This allowed us to prove Brezin-Hikami's conjecture, and to find some explicit form for $n=3$, and conjecture some explicit form in terms of Bessel polynomials for the general case.

For $\beta=2$, we have simplified our recursion (computed the residues). The same method seems to be applicable for higher $\beta$, but we have not done it in this article.

\medskip

We have obtained many new forms of angular integrals, but unfortunately, this does not seem to be the end of the story. Our expressions are still not explicit enough to be useful for computing matrix integrals.
The form of our expressions, strongly suggest that the kernel determinantal formulae \cite{Mehtabook} in the $\beta=1$ case, could be replaced by hyperdeterminantal formulae for higher $\beta$, but this is still to be understood.
The best thing, would be to get expressions with enough structure to generalize the method over integration of matrix variables of Mehta \cite{Mehta1}.

\section*{Acknowledgments}
We would like to thank G. Akemann, E. Brezin, P. Desrosiers, S. Hikami, A. Prats-Ferrer, J.B. Zuber, for useful and fruitful discussions on this subject.
This work is partly supported by the Enigma European network MRT-CT-2004-5652, by the ANR project G\'eom\'etrie et int\'egrabilit\'e en physique math\'ematique ANR-05-BLAN-0029-01, by the Enrage European network MRTN-CT-2004-005616,
by the European Science Foundation through the Misgam program,
by the French and Japaneese governments through PAI Sakurav, by the Quebec government with the FQRNT.

\setcounter{section}{0}

\appendix{Polynomials of $\tau$}\label{applemmaPoltau}

\bl\label{Lemma1Pntau}
Let
\beq
P_n(X,Y) = x_1x_2\dots x_{n-1}x_n\,\, y_{n+1} y_{n+2}\dots y_{2n-1} y_{2n}
+  y_1y_2\dots y_{n-1}y_n\,\, x_{n+1} x_{n+2}\dots x_{2n-1} x_{2n}
\eeq
We prove that $P_n$ is a polynomial of degree $n$ in the $\tau$'s, where $\tau_{i,2n+1}=-{1\over 2} x_i y_i$ and $\tau_{i,j}=-{1\over 2}(x_i-x_j)(y_i-y_j)$, with integer coefficients:
\beq
P_n \in {\mathbb Z}[\tau]
\eeq
\el
\proof{
It clearly holds for $n=0$ and $n=1$.
Indeed the $n=1$ case reads:
\beq
x_1 y_2 + x_2 y_1 = x_1 y_1 + x_2 y_2 - (x_1-x_2)(y_1-y_2) = 2(\tau_{1,2}-\tau_{1,3}-\tau_{2,3})
\eeq

\medskip
Assume that the lemma holds up to $n-1$, and let us prove it for $n$.
In the following $A\equiv B$ means that $A-B$ is a polynomial in the $\tau$'s. 
\bea
&& P_n \cr
&=& x_1x_2\dots x_{n-1}x_n\,\, y_{n+1} y_{n+2}\dots y_{2n-1} y_{2n} \cr
&& +  y_1y_2\dots y_{n-1}y_n\,\, x_{n+1} x_{n+2}\dots x_{2n-1} x_{2n} \cr
&=& (x_1 y_{n+1} + x_{n+1} y_1) (x_2\dots x_{n-1}x_n\,\, y_{n+2}\dots y_{2n-1} y_{2n} \cr
&& +  y_2\dots y_{n-1}y_n\,\,  x_{n+2}\dots x_{2n-1} x_{2n}) \cr
&& -  x_2\dots x_{n-1}x_n x_{n+1}\,\, y_{n+2}\dots y_{2n-1} y_{2n} y_1 \cr  
&& - y_2\dots y_{n-1}y_n y_{n+1}\,\,   x_{n+2}\dots x_{2n-1} x_{2n} x_1 \cr
&\equiv &  -  x_2\dots x_{n-1}x_n x_{n+1}\,\, y_{n+2}\dots y_{2n-1} y_{2n} y_1 \cr
&& - y_2\dots y_{n-1}y_n y_{n+1}\,\,   x_{n+2}\dots x_{2n-1} x_{2n} x_1 \cr
\eea
and then:
\bea
&& P_n \cr
&\equiv &  -  x_2\dots x_{n-1}x_n x_{n+1}\,\, y_{n+2}\dots y_{2n-1} y_{2n} y_1 \cr
&& - y_2\dots y_{n-1}y_n y_{n+1}\,\,   x_{n+2}\dots x_{2n-1} x_{2n} x_1 \cr
&\equiv &  - (x_2 y_1 + x_1 y_2) ( x_3\dots x_{n-1}x_n x_{n+1}\,\, y_{n+2}\dots y_{2n-1} y_{2n} \cr
&& + y_3\dots y_{n-1}y_n y_{n+1}\,\, x_{n+2}\dots x_{2n-1} x_{2n}) \cr
&&  + x_{n+1} x_1 x_3\dots x_{n-1}x_n \,\, y_2 y_{n+2}\dots  y_{2n} + y_{n+1} y_1 y_3\dots y_n \,\, x_2 x_{n+2}\dots  x_{2n} \cr
\eea

Repeating the same operation recursively we obtain $\forall k$:
\bea
P_n 
&\equiv & x_{n+1}\dots x_{n+k}\, x_1\,  x_{k+2}\dots  x_{n}\,\, y_2 \dots y_{k+1} \,  y_{k+n+1}\dots y_{2n} \cr
&& + y_{n+1}\dots y_{n+k}\, y_1\,  y_{k+2}\dots  y_{n}\,\, x_2 \dots x_{k+1} \,  x_{k+n+1}\dots x_{2n}  \cr
\eea

In particular for $k=n-2$ we find:
\bea
P_n 
&\equiv & x_{n+1}\dots x_{2n-2}\, x_1\, x_{n}\,\, y_2 \dots y_{n-1} \,  y_{2n-1} y_{2n}  \cr
&& + y_{n+1}\dots y_{2n-2}\, y_1\,  y_{n}\,\, x_2 \dots x_{n-1} \,  x_{2n-1} x_{2n}  \cr
&\equiv & (x_1 y_{2n-1}+ x_{2n-1} y_1) (x_{n+1}\dots x_{2n-2}\,\, x_{n}\,\, y_2 \dots y_{n-1} \,  y_{2n}  \cr
&& + y_{n+1}\dots y_{2n-2}\, \,  y_{n}\,\, x_2 \dots x_{n-1} \,   x_{2n} ) \cr
& & - x_{n}\dots x_{2n-1}\,\,  y_1 y_2 \dots y_{n-1} \,  y_{2n} 
- y_{n}\dots y_{2n-1}\, \,  \,\,x_1 x_2 \dots x_{n-1} \,   x_{2n} ) \cr
\eea
and we repeat the same operations:
\bea
P_n 
&\equiv & - x_{n}\dots x_{2n-1}\,\,  y_1 y_2 \dots y_{n-1} \,  y_{2n} 
- y_{n}\dots y_{2n-1}\, \,  \,\,x_1 x_2 \dots x_{n-1} \,   x_{2n} ) \cr
&\equiv & - (x_n y_1+x_1 y_n) ( x_{n+1}\dots x_{2n-1}\,\,  y_2 \dots y_{n-1} \,  y_{2n} \cr
&& + y_{n+1}\dots y_{2n-1}\, \,  x_2 \dots x_{n-1} \,   x_{2n} ) \cr
&& +  x_1 x_{n+1}\dots x_{2n-1}\,\,  y_2 \dots y_{n} \,  y_{2n} 
+ y_1 y_{n+1}\dots y_{2n-1}\, \,  x_2 \dots x_{n} \,   x_{2n}  \cr
\eea
and once more
\bea
P_n 
&\equiv&   x_1 x_{n+1}\dots x_{2n-1}\,\,  y_2 \dots y_{n} \,  y_{2n} 
+ y_1 y_{n+1}\dots y_{2n-1}\, \,  x_2 \dots x_{n} \,   x_{2n}  \cr
&\equiv&  (x_1 y_{2n}+x_{2n} y_1) ( x_{n+1}\dots x_{2n-1}\,\,  y_2 \dots y_{n}  
+ y_{n+1}\dots y_{2n-1}\, \,  x_2 \dots x_{n} )  \cr
&& -  x_{n+1}\dots x_{2n}\,\, y_1  y_2 \dots y_{n}  - y_{n+1}\dots y_{2n}\, \,  x_1 \dots x_{n}   \cr
\eea
Therefore $P_n\equiv -P_n$, i.e. $P_n$ is a polynomial in the $\tau$'s.

}

\bl\label{Lemma2Pntau}
Let
\beq
P_{\alpha,\beta}(X,Y) = \prod_{i=1}^n x_i^{\alpha_i} y_i^{\beta_i} + 
\prod_{i=1}^n x_i^{\beta_i} y_i^{\alpha_i} 
\virg \sum_i \alpha_i=\sum_i \beta_i=d
\eeq
We prove that $P_{\alpha,\beta}$ is a polynomial of degree $d$ in the $\tau$'s, with integer coefficients:
\beq
P_n \in {\mathbb Z}[\tau]
\eeq

\el

\proof{
We proceed by recursion on the total degree $d=\sum_i \alpha_i=\sum_i \beta_i$.
The lemma clearly holds for $d=0$ and $d=1$.

\medskip
Assume $d\geq 2$ and that the lemma holds up to $d-1$. We will prove it for $d$.

\medskip

$\bullet$ if there exists $i$ such that $\alpha_i \beta_i>0$, then $P_{\alpha,\beta}$ is the factor of $\tau_{i,n+1}$ times $P_{\alpha',\beta'}$ of smaller degree, and from the recursion hypothesis it holds.

$\bullet$ assume that $\forall i, \alpha_i \beta_i=0$. Since $d\geq 2$, there must exist some $i$ such that $\alpha_i\geq 1$ and some $j$ such that $\beta_j\geq 1$.
Let us choose $k$ and $l$ such that:
\beq
\alpha_k = \mathop{{\rm max}}_i \{ \alpha_i\} \geq 1
\virg
\beta_l = \mathop{{\rm max}}_i \{ \beta_i\} \geq 1
\eeq
We have $k\neq l$.

If $\alpha_k=\beta_l=1$, then we can apply lemma\ref{Lemma1Pntau}, and thus the lemma is proved for that case.

$\bullet$ therefore, we now assume that $\alpha_k \beta_l\geq 2$, and with no loss of generality we may assume that $\alpha_k\geq 2$.
We write:
\bea
&& P_n \cr
&=& x_1^{\alpha_1} \dots x_n^{\alpha_n}\, y_1^{\beta_1} \dots y_n^{\beta_n} + x_1^{\beta_1} \dots x_n^{\beta_n}\, y_1^{\alpha_1} \dots y_n^{\alpha_n} \cr
&=& (x_k y_l + x_l y_k) ( x_1^{\alpha_1} \dots x_k^{\alpha_k-1} \dots x_n^{\alpha_n}\, y_1^{\beta_1} \dots y_l^{\beta_l-1} \dots y_n^{\beta_n} \cr
&& + y_1^{\alpha_1} \dots y_k^{\alpha_k-1} \dots y_n^{\alpha_n}\, x_1^{\beta_1} \dots x_l^{\beta_l-1} \dots x_n^{\beta_n} ) \cr
&& -  x_1^{\alpha_1} \dots x_k^{\alpha_k-1} x_l^{\alpha_l+1}  \dots x_n^{\alpha_n}\, y_1^{\beta_1} \dots y_l^{\beta_l-1} y_k^{\beta_k+1}  \dots y_n^{\beta_n} \cr
&&  -  y_1^{\alpha_1} \dots y_k^{\alpha_k-1} y_l^{\alpha_l+1}  \dots y_n^{\alpha_n}\, x_1^{\beta_1} \dots x_l^{\beta_l-1} x_k^{\beta_k+1}  \dots x_n^{\beta_n} \cr
&\equiv& -x_k y_k\,\, (  x_1^{\alpha_1} \dots x_k^{\alpha_k-2} x_l^{\alpha_l+1}  \dots x_n^{\alpha_n}\, y_1^{\beta_1} \dots y_l^{\beta_l-1} y_k^{\beta_k}  \dots y_n^{\beta_n} \cr
&&  +  y_1^{\alpha_1} \dots y_k^{\alpha_k-2} y_l^{\alpha_l+1}  \dots y_n^{\alpha_n}\, x_1^{\beta_1} \dots x_l^{\beta_l-1} x_k^{\beta_k}  \dots x_n^{\beta_n} ) \cr
\eea
From the recursion hypothesis, the RHS is a polynomial in the $\tau$'s, and thus we have proved the lemma.

}

\bl\label{LemmaPsymtau}

Let $P$ be a polynomial of $2n+2$ variables $x_1,\dots,x_n,x_{n+1},y_1,\dots,y_n,y_{n+1}$, with the following properties:
\begin{itemize}
\item $P$ is invariant by translations $\forall i,\,\,\, x_i\to x_i+\delta x$, $y_i\to y_i+\delta y$,

\item $P$ is invariant under $\forall i, \,\,\, x_i\to \l x_i$, $y_i\to {1\over \l}y_i$ ,

\item $P$ is symmetric in the exchange $X\leftrightarrow Y$,


\end{itemize}
Then $P$ is a polynomial of the $\tau_{i,j}$'s.

\el

\proof{
Because of invariance by translation, we can always assume that $x_{n+1}=y_{n+1}=0$.
Then, the other properties imply that $P$ is a linear combination of monomials of the type $P_{\alpha,\beta}(X,Y)$ of Lemma.\ref{Lemma2Pntau}.

}

\appendix{Loop equations}
\label{apploopeq}

Loop equations for matrix models have been studied for a long time \cite{Migdal}. Loop equations (sometimes called Ward identities or Schwinger-Dyson equations) for $\beta$ ensembles can be found for instance in \cite{eynbeta, WZ1, Jevicki1, Jevicki2, Jevicki3, Cheynbeta}.
Here, we summarize the method.

\bigskip

$\bullet$ 1-matrix model in eigenvalue representation, for arbitrary $\beta$:

Consider the integral:
\beq
Z=\int d\l_1 \dots d\l_n \,\, \ee{-\sum_i V(\l_i)}\,\,\, \prod_{i<j} (\l_j-\l_i)^{2\beta}
\eeq
If we make an infinitesimal local change of variable $\l_i\to \l_i+\epsilon \l_i^k + O(\epsilon^2)$, we find:
\bea
Z
&=&(1+O(\epsilon^2))\,\int d\l_1 \dots d\l_n \prod_i (1+\epsilon k\l_i^{k-1}) \,\, \ee{-\sum_i V(\l_i)} \prod_i (1-\epsilon \l_i^k V'(\l_i))\,\,\,\cr
&&  \prod_{i<j} (\l_j-\l_i)^{2\beta} \,\, \prod_{i< j}(1+2\beta\epsilon {\l_i^k-\l_j^k\over \l_i-\l_j})  \cr
\eea
i.e., by considering the term linear in $\epsilon$:
\beq
 0 = \left< \sum_i k\l_i^{k-1} + \beta\sum_{l=0}^{k-1}\sum_{i\neq j} \l_i^l\l_j^{k-1-l} -  \sum_i \l_i^k V'(\l_i) \right>
\eeq
It can be written collectively by summing over $k$ with $1/x^{k+1}$, this is equivalent to consider 
a local change of variable $\l_i\to \l_i+{\epsilon \over x-\l_i} + O(\epsilon^2)$, and we write $\om(x) = \sum_i {1\over x-\l_i}$:
\beq
 0 = \left< -\om'(x) + \beta (\om(x)^2+\om'(x)) -  \sum_i {V'(\l_i)\over x-\l_i} \right>
\eeq
i.e.
\beq\label{loopeqbeta1app}
0=-\left< \Tr{V'(M)\over x-M}\right>
+\beta \left< \Tr{1\over x-M}\,\Tr{1\over x-M}\right>
+(\beta-1)\,{\partial\over\partial x}\, \left< \Tr{1\over x-M}\right>
\eeq

$\bullet$ 2-matrix model for $\beta=1/2,1,2$:

Similarly if we consider a 2-matrix model:
\beq
Z=\int dM_1 dM_2 \,\, \ee{-\Tr(V_1(M_1)+V_2(M_2)-M_1 M_2)}
\eeq
again we make a local change of variable $M_1\to M_1 + \epsilon {1\over x-M_1} A + O(\epsilon^2)$.
The Jacobian of this change of variable is computed as a split rule (cf \cite{}), it can be computed for each of the 3 ensembles $\beta=1/2,1,2$ and is worth:
\beq
dM_1 \to dM_1 \left(1+\epsilon \beta \Tr {1\over x-M_1}\Tr A\,{1\over x-M_1} + \epsilon (\beta-1) \,{\partial\over\partial x}\,  \Tr A\,{1\over x-M_1}+ O(\epsilon^2) \right)
\eeq
Thus we find:
\bea\label{loopeqbeta2app}
0&=& -\left< \Tr(V'_1(M_1)-M_2)\,{1\over x-M_1}\right>
+\beta \left< \Tr{1\over x-M_1}\,\Tr A\,{1\over x-M_1}\right> \cr
&& +(\beta-1)\,{\partial\over\partial x}\, \left< \Tr A\,{1\over x-M_1}\right>
\eea

$\bullet$ Without a definition of a 2-matrix integral for arbitrary $\beta$, it is not possible to find the loop equation for any $\beta$.
However, we see that equation \eq{loopeqbeta2app} is valid for the 2-matrix model for $\beta=1/2,1,2$, and is valid for the 1-matrix eigenvalue model for any $\beta$.
Therefore it is natural to take it as a definition of the 2-matrix model for arbitrary $\beta$.

\appendix{Proof $n=3$}\label{appproofn3}

%



\subsection{BESSEL ZOOLOGY}

\bigskip

We define the functions%
\begin{equation}
Q_{\beta }\left( x\right) =\sum_{l=0}^{\infty }\left( -\right) ^{l}\ \frac{%
\Gamma \left( \beta +l\right) }{\Gamma \left( \beta -l\right) }\ \frac{2^{-l}%
}{l!}\ x^{\beta -l}  \label{bergn31}
\end{equation}%
We have%
\begin{eqnarray}
2x^{2}\ Q_{\beta }^{\prime }\left( x\right) -2\beta \ x\ Q_{\beta }\left(
x\right) 
&=&-\sum_{l=1}^{\infty }\left( -\right) ^{l}\ \frac{\Gamma \left(
\beta +l\right) }{\Gamma \left( \beta -l\right) }\ \frac{2^{-l+1}}{\left(
l-1\right) }\ x^{\beta -l+1}  \label{bergn32a} \cr
x^{2}\ Q_{\beta }^{"}\left( x\right) -2\beta \ x\ Q_{\beta }^{\prime }\left(
x\right) +2\beta \ Q_{\beta }\left( x\right) 
&=&-\sum_{l=0}^{\infty }\left(
-\right) ^{l}\ \frac{\Gamma \left( \beta +l+1\right) }{\Gamma \left( \beta
-l-1\right) }\ \frac{2^{-l}}{l!}\ x^{\beta -l}  \label{bergn32b}
\end{eqnarray}%
Changing $l$ into$\ l+1$ in \ref{bergn32a} we obtain the differential equation%
\begin{equation}
x^{2}\ Q_{\beta }^{"}\left( x\right) -2x\left( \beta -x\right) \ Q_{\beta
}^{\prime }\left( x\right) +2\beta \left( 1-x\right) \ Q_{\beta }\left(
x\right) =0  \label{(3)}
\end{equation}

\bigskip

We define the functions%
\begin{eqnarray}
Q_{\beta ,k}\left( x\right) &=&\left( -2\right) ^{k}\ x^{\beta -k}\ \left(
x^{2}\frac{d}{dx}\right) ^{k}\ \left( \frac{Q_{\beta }\left( x\right) }{%
x^{\beta }}\right)  \label{(4a)} \\
Q_{\beta ,0}\left( x\right) &=&Q_{\beta }\left( x\right)  \label{(4b)}
\end{eqnarray}%
that is%
\begin{equation}
Q_{\beta ,k}\left( x\right) =\sum_{l=0}^{\infty }\left( -\right) ^{l+k}\ 
\frac{\Gamma \left( \beta +l+k\right) }{\Gamma \left( \beta -l-k\right) }\ 
\frac{2^{-l}}{l!}\ x^{\beta -l-k}  \label{(5)}
\end{equation}%
From \ref{(4a)} we obtain the recurrence%
\begin{equation}
Q_{\beta ,k}\left( x\right) =2\left( \beta -k+1\right) \ Q_{\beta
,k-1}\left( x\right) -2x\ Q_{\beta ,k-1}^{\prime }\left( x\right)  \label{(6)}
\end{equation}%
For instance%
\begin{equation}
Q_{\beta ,1}\left( x\right) =2\beta \ Q_{\beta ,0}\left( x\right) -2x\
Q_{\beta ,0}^{\prime }\left( x\right)  \label{(7)}
\end{equation}%
and%
\begin{eqnarray}
Q_{\beta ,2}\left( x\right) &=&2\left( \beta -1\right) \ Q_{\beta ,1}\left(
x\right) -2x\ Q_{\beta ,1}^{\prime }\left( x\right)  \label{(8a)} \\
Q_{\beta ,2}\left( x\right) &=&4x^{2}\ Q_{\beta ,0}^{"}\left( x\right)
-8x\left( \beta -1\right) \ Q_{\beta ,0}^{\prime }\left( x\right) +4\beta
\left( \beta -1\right) \ Q_{\beta ,0}\left( x\right)  \label{(8b)}
\end{eqnarray}%
Thus, we have, from \ref{(7)} and \ref{(8b)}, the expressions $Q_{\beta ,0}^{\prime
}\left( x\right) $ and $Q_{\beta ,0}^{"}\left( x\right) $ in terms of $%
Q_{\beta ,1}\left( x\right) $ and $Q_{\beta ,2}\left( x\right) $. Equation
\ref{(3)} becomes%
\begin{equation}
Q_{\beta ,2}\left( x\right) +4\left( 1-x\right) \ Q_{\beta ,1}\left(
x\right) -4\beta \left( \beta -1\right) \ Q_{\beta ,0}\left( x\right) =0 
\label{(9)}
\end{equation}

\bigskip

From \ref{(9)}, by derivatives $\frac{d}{dx}\ $and recurrence, it is easy to show
that%
\begin{equation}
Q_{\beta ,k+2}\left( x\right) +4\left( k+1-x\right) \ Q_{\beta ,k+1}\left(
x\right) -4\left[ \beta \left( \beta -1\right) -k\left( k+1\right) \right] \
Q_{\beta ,k}\left( x\right) =0  \label{(10)}
\end{equation}

\bigskip

\subsection{CALOGERO N=3}

\bigskip

We consider the Calogero differential operator%
\begin{equation}
H_{\rm Calogero}=\sum_{i=1}^{3}\frac{d^{2}}{dx_{i}^{2}}+2\beta \sum_{i<j}\frac{1}{%
x_{i}-x_{j}}\left( \frac{d}{dx_{i}}-\frac{d}{dx_{j}}\right)   \label{(20)}
\end{equation}%
and we look for solutions 
\begin{equation}
H_{\rm Calogero}\ \Phi \left( x_{i},\ y_{i}\right) =\left( \sum_{i}y_{i}^{2}\right)
\ \ \Phi \left( x_{i},\ y_{i}\right)   \label{(21)}
\end{equation}%
where the solutions $\Phi \left( x_{i},\ y_{i}\right) $have a certain number
of symmetry properties described somewhere else.\ We write%
\begin{equation}
\Phi \left( x_{i},\ y_{i}\right) =f\left( x_{i},\ y_{i}\right) \
e^{\sum_{i=1}^{3}x_{i}y_{i}}  \label{(22)}
\end{equation}%
so that the equations \ref{(20)} and \ref{(21)} become%
\begin{equation}
D=\sum_{i=1}^{3}\frac{d^{2}}{dx_{i}^{2}}+2\beta \sum_{i<j}\frac{1}{%
x_{i}-x_{j}}\left( \frac{d}{dx_{i}}-\frac{d}{dx_{j}}+y_{i}-y_{j}\right)
+2\sum_{i=1}^{3}y_{i}\frac{d}{dx_{i}}  \label{(23a)}
\end{equation}%
and%
\begin{equation}
D\ f\left( x_{i},\ y_{i}\right) =0  \label{(23b)}
\end{equation}

\bigskip

Let us introduce the variables%
\begin{eqnarray}
a &=&\frac{1}{2}\left( x_{1}-x_{2}\right) \ \left( y_{1}-y_{2}\right)
=\left( x_{1}-x_{2}\right) \ Y_{12}  \label{(24a)} \\
b &=&\frac{1}{2}\left( x_{1}-x_{3}\right) \ \left( y_{1}-y_{3}\right)
=\left( x_{1}-x_{3}\right) \ Y_{13}  \label{(24b)} \\
c &=&\frac{1}{2}\left( x_{2}-x_{3}\right) \ \left( y_{2}-y_{3}\right)
=\left( x_{2}-x_{3}\right) \ Y_{23}  \label{(24c)}
\end{eqnarray}%
where 
\begin{equation}
Y_{ij}=-Y_{ji}  \label{(25)}
\end{equation}%
and we look for solutions of the type $f\left( a,b,c\right) .$ We have%
\begin{eqnarray}
\frac{df}{dx_{1}} &=&Y_{12}\ f_{a}^{\prime }+Y_{13}f_{b}^{\prime } 
\label{(26a)} \\
\frac{df}{dx_{2}} &=&Y_{21}\ f_{a}^{\prime }+Y_{23}f_{c}^{\prime } 
\label{(26b)} \\
\frac{df}{dx_{3}} &=&Y_{31}\ f_{b}^{\prime }+Y_{32}f_{c}^{\prime } 
\label{(26c)}
\end{eqnarray}%
and%
\begin{eqnarray}
\frac{d^{2}f}{dx_{1}^{2}}
&=&Y_{12}^{2}f_{a^{2}}^{"}+2Y_{12}Y_{13}f_{ab}^{"}+Y_{13}^{2}f_{b^{2}}^{"} 
\label{(27a)} \\
\frac{d^{2}f}{dx_{2}^{2}}
&=&Y_{21}^{2}f_{a^{2}}^{"}+2Y_{21}Y_{23}f_{ac}^{"}+Y_{23}^{2}f_{c^{2}}^{"} 
\label{(27b)} \\
\frac{d^{2}f}{dx_{3}^{2}}
&=&Y_{31}^{2}f_{b^{2}}^{"}+2Y_{31}Y_{32}f_{bc}^{"}+Y_{32}^{2}f_{c^{2}}^{"} 
\label{(27c)}
\end{eqnarray}%
The equations \ref{(23a)}\ and \ref{(23b)}\ become%
\begin{eqnarray}
D\ f\left( a,b,c\right) &=&D_{1}\ f\left( a,b,c\right) +D_{2}\ f\left(
a,b,c\right) =0  \label{(28a)} \\
D_{1}\ f\left( a,b,c\right) &=&Y_{12}^{2}\left[ 2f_{a^{2}}^{"}+4(\frac{\beta 
}{a}+1)\ f_{a}^{\prime }+4\frac{\beta }{a}\ f\right] +circ.perm. 
\label{(28b)} \\
D_{2}\ f\left( a,b,c\right) &=&2Y_{12}Y_{13}\left[ f_{ab}^{"}+\frac{\beta }{a%
}\ f_{b}^{\prime }+\frac{\beta }{b}\ f_{a}^{\prime }\right] +circ.perm. 
\label{(28c)}
\end{eqnarray}

\bigskip

We now try the functions%
\begin{equation}
f\left( a,b,c\right) =\frac{Q_{\beta ,k}\left( a\right) }{a^{2\beta }}\ 
\frac{Q_{\beta ,k}\left( b\right) }{b^{2\beta }}\ \frac{Q_{\beta ,k}\left(
c\right) }{c^{2\beta }}=\frac{\left\{ k,k,k\right\} }{\left( abc\right)
^{2\beta }}  \label{(29)}
\end{equation}%
where $Q_{\beta ,k}\left( x\right) $ are defined in (\ref{(4a)}-\ref{(4b)}) and \ref{(5)}. We
consider%
\begin{equation}
f\left( a\right) =\frac{Q_{\beta ,k}\left( a\right) }{a^{2\beta }} 
\label{(30)}
\end{equation}%
we have%
\begin{equation}
2\ f_{a}^{\prime }+2\frac{\beta }{a}\ f=\frac{2a\ Q_{\beta ,k}^{\prime
}\left( a\right) -2\beta \ Q_{\beta ,k}\left( a\right) }{a^{^{2\beta +1}}}=-%
\frac{2k\ Q_{\beta ,k}\left( a\right) +Q_{\beta ,k+1}\left( a\right) \ }{%
a^{^{2\beta +1}}}  \label{(31)}
\end{equation}%
By derivation we obtain%
\begin{equation}
2f_{a^{2}}^{"}+2\frac{\beta }{a}\ f_{a}^{\prime }-2\frac{\beta }{a^{2}}\ f=%
\frac{1}{a^{^{2\beta +2}}}\left[ 
\begin{array}{c}
\frac{1}{2}Q_{\beta ,k+2}\left( a\right) +\left( \beta +2k+2\right) Q_{\beta
,k+1}\left( a\right)  \\ 
+2k\left( \beta +k+1\right) Q_{\beta ,k}\left( a\right) 
\end{array}%
\right]   \label{(32)}
\end{equation}%
so that%
\begin{eqnarray}
2f_{a^{2}}^{"}+4\frac{\beta }{a}\ f_{a}^{\prime }+\frac{2\beta \left( \beta
-1\right) }{a^{2}}\ f &=&\frac{1}{a^{^{2\beta +2}}}\left[ 
\begin{array}{c}
\frac{1}{2}Q_{\beta ,k+2}\left( a\right) +2\left( k+1\right) Q_{\beta
,k+1}\left( a\right)  \\ 
+2k\left( k+1\right) Q_{\beta ,k}\left( a\right) 
\end{array}%
\right]   \label{(33a)} \\
2f_{a^{2}}^{"}+4\frac{\beta }{a}\ f_{a}^{\prime } &=&\frac{1}{a^{^{2\beta
+2}}}\left[ 
\begin{array}{c}
\frac{1}{2}Q_{\beta ,k+2}\left( a\right) +2\left( k+1\right) Q_{\beta
,k+1}\left( a\right)  \\ 
-2\left[ \beta \left( \beta -1\right) -k\left( k+1\right) \right] Q_{\beta
,k}\left( a\right) 
\end{array}%
\right]   \label{(33b)}
\end{eqnarray}%
Finally we obtain%
\begin{equation}
2f_{a^{2}}^{"}+4(\frac{\beta }{a}+1)\ f_{a}^{\prime }+4\frac{\beta }{a}\ f=%
\frac{1}{2a^{^{2\beta +2}}}\left[ 
\begin{array}{c}
Q_{\beta ,k+2}\left( a\right) +4\left( k+1-a\right) Q_{\beta ,k+1}\left(
a\right)  \\ 
-4\left[ \beta \left( \beta -1\right) -k\left( k+1-2a\right) \right]
Q_{\beta ,k}\left( a\right) 
\end{array}%
\right]   \label{(34)}
\end{equation}%
Now, we use the recurrence relation \ref{(10)} and get the simple result%
\begin{equation}
2f_{a^{2}}^{"}+4(\frac{\beta }{a}+1)\ f_{a}^{\prime }+4\frac{\beta }{a}\ f=-%
\frac{4k}{a^{^{2\beta +1}}}\ Q_{\beta ,k}\left( a\right)   \label{(35)}
\end{equation}%
We just proved that%
\begin{equation}
D_{1}\ \frac{\left\{ k,k,k\right\} }{\left( abc\right) ^{2\beta }}=-4k\left( 
\frac{Y_{12}^{2}}{a}+\frac{Y_{31}^{2}}{b}+\frac{Y_{23}^{2}}{c}\right) \ 
\frac{\left\{ k,k,k\right\} }{\left( abc\right) ^{2\beta }}  \label{(36)}
\end{equation}%
We further transform the result \ref{(36)}. We have%
\begin{eqnarray}
\frac{Y_{12}^{2}}{a} &=&\frac{Y_{12}}{\left( x_{1}-x_{2}\right) }=\frac{%
Y_{12}}{\Delta \left( x\right) }\left( x_{1}-x_{3}\right) \ \left(
x_{2}-x_{3}\right)   \label{(37a)} \\
\frac{Y_{12}^{2}}{a} &=&\frac{Y_{12}}{\Delta \left( x\right) }\left[
F-\left( x_{1}-x_{2}\right) ^{2}\right]   \label{(37b)}
\end{eqnarray}%
where%
\begin{eqnarray}
\Delta \left( x\right)  &=&\left( x_{1}-x_{2}\right) \ \left(
x_{1}-x_{3}\right) \ \left( x_{2}-x_{3}\right)   \label{(38a)} \\
F &=&x_{1}^{2}+x_{2}^{2}+x_{3}^{2}-x_{1}x_{2}-x_{1}x_{3}-x_{2}x_{3} 
\label{(38b)}
\end{eqnarray}%
By circular permutation we also have%
\begin{eqnarray}
\frac{Y_{31}^{2}}{b} &=&\frac{Y_{31}}{\Delta \left( x\right) }\left[
F-\left( x_{3}-x_{1}\right) ^{2}\right]   \label{(39a)} \\
\frac{Y_{23}^{2}}{c} &=&\frac{Y_{23}}{\Delta \left( x\right) }\left[
F-\left( x_{2}-x_{3}\right) ^{2}\right]   \label{(39b)}
\end{eqnarray}%
We note that in \ref{(36)} the quantity $F\ $disappear since%
\begin{equation}
Y_{12}+Y_{23}+Y_{31}=0  \label{(40)}
\end{equation}%
We may write now%
\begin{equation}
D_{1}\ \frac{\left\{ k,k,k\right\} }{\left( abc\right) ^{2\beta }}=\frac{4k}{%
\Delta \left( x\right) }\left[ \left( x_{1}-x_{2}\right) \ a+circ.perm.%
\right] \ \frac{\left\{ k,k,k\right\} }{\left( abc\right) ^{2\beta }} 
\label{(41)}
\end{equation}

\bigskip

We now consider 
\begin{equation}
D_{2}\ \frac{\left\{ k,k,k\right\} }{\left( abc\right) ^{2\beta }} 
\label{(42)}
\end{equation}%
Using%
\begin{equation}
f(a,b,c)=f(a)\ f(b)\ f(c)  \label{(43)}
\end{equation}%
we have%
\begin{equation}
f_{ab}^{"}(a,b,c)+\frac{\beta }{a}\ f_{b}^{\prime }(a,b,c)+\frac{\beta }{b}\
f_{a}^{\prime }(a,b,c)=\left[ 
\begin{array}{c}
\left( f_{a}^{\prime }(a)+\frac{\beta f(a)}{a}\right) \left( f_{b}^{\prime
}(b)+\frac{\beta f(b)}{b}\right) \\ 
-\frac{\beta ^{2}f(a)\ f(b)}{ab}%
\end{array}%
\right] \ f(c)  \label{(44)}
\end{equation}%
but in \ref{(28c)}\ the term $\frac{\beta ^{2}}{ab}$ $f(a)\ f(b)\ f(c)\ $%
disappears since%
\begin{equation}
\frac{Y_{12}Y_{13}}{ab}+circ.perm.=\frac{x_{2}-x_{3}}{\Delta \left( x\right) 
}+circ.perm.=0  \label{(45)}
\end{equation}%
Then, from \ref{(31)} we get%
\begin{equation}
=\left[ 
\begin{array}{c}
2Y_{12}Y_{13}\frac{\left[ k\ Q_{\beta ,k}\left( a\right) +\frac{1}{2}%
Q_{\beta ,k+1}\left( a\right) \right] }{a^{2\beta +1}}\ \frac{\left[ k\
Q_{\beta ,k}\left( b\right) +\frac{1}{2}Q_{\beta ,k+1}\left( b\right) \right]
}{b^{2\beta +1}}\ \ \frac{Q_{\beta ,k}\left( c\right) }{c^{2\beta }} \\ 
+circ.perm.%
\end{array}%
\right]  \label{(46)}
\end{equation}%
Again, the term containing $Q_{\beta ,k}\left( a\right) \ Q_{\beta ,k}\left(
b\right) \ \ Q_{\beta ,k}\left( c\right) $ disappears by \ref{(45)}. We write%
\bea
D_{2}\ \frac{\left\{ k,k,k\right\} }{\left( abc\right) ^{2\beta }}
&=& \frac{1}{%
\Delta \left( x\right) }\frac{1}{\left( abc\right) ^{2\beta }}\Big[ \left(
x_{2}-x_{3}\right) \ \left[ 
\begin{array}{c}
\frac{1}{2}\left\{ k+1,k+1,k\right\} \\ 
+k\ \left\{ k+1,k,k\right\} +k\ \left\{ k,k+1,k\right\}%
\end{array}%
\right] \cr 
&& \qquad +circ.perm.\Big]  \label{(47)}
\eea%
We note that%
\begin{equation}
\left( x_{2}-x_{3}\right) \left[ \left\{ k+1,k,k\right\} +\left\{
k,k+1,k\right\} +\left\{ k,k,k+1\right\} \right] +circ.perm.=0  \label{(48)}
\end{equation}%
so that%
\begin{equation}
D_{2}\ \frac{\left\{ k,k,k\right\} }{\left( abc\right) ^{2\beta }}=\frac{1}{%
\Delta \left( x\right) }\frac{1}{\left( abc\right) ^{2\beta }}\left[ \left(
x_{2}-x_{3}\right) \ \left[ 
\begin{array}{c}
\frac{1}{2}\left\{ k+1,k+1,k\right\} \\ 
-k\ \left\{ k,k,k+1\right\}%
\end{array}%
\right] +circ.perm.\right]  \label{(49)}
\end{equation}

\bigskip

We now collect$\ D_{1}$ and$\ D_{2}$. From \ref{(28a)}, \ref{(29)}, \ref{(41)} and \ref{(49)} we
obtain%
\begin{equation}
D\ \frac{\left\{ k,k,k\right\} }{\left( abc\right) ^{2\beta }}=\frac{1}{%
\Delta \left( x\right) }\frac{1}{\left( abc\right) ^{2\beta }}\left[ 
\begin{array}{c}
\left( x_{2}-x_{3}\right) \left[ 
\begin{array}{c}
4kc\left\{ k,k,k\right\} +\frac{1}{2}\left\{ k+1,k+1,k\right\} \\ 
-k\ \left\{ k,k,k+1\right\}%
\end{array}%
\right] \\ 
+circ.perm.%
\end{array}%
\right]  \label{(50)}
\end{equation}%
Again,%
\begin{equation}
\left( x_{2}-x_{3}\right) \left\{ k,k,k\right\} +circ.perm.=0  \label{(51)}
\end{equation}%
so\ that 
\begin{equation}
D\ \frac{\left\{ k,k,k\right\} }{\left( abc\right) ^{2\beta }}=\frac{1}{%
\Delta \left( x\right) }\frac{1}{\left( abc\right) ^{2\beta }}\left[ 
\begin{array}{c}
\left( x_{2}-x_{3}\right) \left[ 
\begin{array}{c}
4k\left( c-k\right) \left\{ k,k,k\right\} \\ 
+\frac{1}{2}\left\{ k+1,k+1,k\right\} -k\ \left\{ k,k,k+1\right\}%
\end{array}%
\right] \\ 
+circ.perm.%
\end{array}%
\right]  \label{(52)}
\end{equation}%
We now use equation (10) and write%
\bea
D\ \frac{\left\{ k,k,k\right\} }{\left( abc\right) ^{2\beta }}
&=& \frac{1}{%
\Delta \left( x\right) }\frac{1}{\left( abc\right) ^{2\beta }}\left[ 
\begin{array}{c}
\left( x_{2}-x_{3}\right) \big[ \frac{1}{2}\left\{ k+1,k+1,k\right\} \cr
-4k\left( \beta -k\right) \left( \beta +k-1\right) \left\{ k,k,k-1\right\} %
\big] \\ 
+circ.perm.%
\end{array}%
\right]  \label{(53)}
\eea%
Consequently, we obtain the remarquable result%
\bea
&& D\ \left[ \frac{\Gamma \left( \beta -k\right) }{\Gamma \left( \beta
+k\right) }\frac{1}{8^{k}\ k!}\ \ \frac{\left\{ k,k,k\right\} }{\left(
abc\right) ^{2\beta }}\right] \  \cr
&=& \frac{1}{2\Delta \left( x\right) }\frac{1}{%
\left( abc\right) ^{2\beta }}\left[ 
\begin{array}{c}
\left( x_{2}-x_{3}\right) \left[ 
\begin{array}{c}
\frac{\Gamma \left( \beta -k\right) }{\Gamma \left( \beta +k\right) }\frac{1%
}{8^{k}\ k!}\ \ \left\{ k+1,k+1,k\right\} \\ 
-\frac{\Gamma \left( \beta -k+1\right) }{\Gamma \left( \beta +k-1\right) }%
\frac{1}{8^{k-1}\ \left( k-1\right) !}\ \ \left\{ k,k,k-1\right\}%
\end{array}%
\right] \\ 
+circ.perm.%
\end{array}%
\right]  \label{(54)}
\eea%
Clearly enough, we define for $\beta $ not integer%
\begin{equation}
f\left( abc\right) =\sum_{k=0}^{\infty }\frac{\Gamma \left( \beta -k\right) 
}{\Gamma \left( \beta +k\right) }\frac{1}{8^{k}\ k!}\ \ \frac{\left\{
k,k,k\right\} }{\left( abc\right) ^{2\beta }}  \label{(55)}
\end{equation}%
then,%
\begin{equation}
D\ f\left( abc\right) =0+circ.perm.  \label{(56)}
\end{equation}%
Now, if $\beta $ is an integer%
\begin{equation}
Q_{\beta ,k\geq \beta }\left( x\right) =0  \label{(57)}
\end{equation}%
and we define%
\begin{equation}
f\left( abc\right) =\sum_{k=0}^{\beta -1}\frac{\Gamma \left( \beta -k\right) 
}{\Gamma \left( \beta +k\right) }\frac{1}{8^{k}\ k!}\ \ \frac{\left\{
k,k,k\right\} }{\left( abc\right) ^{2\beta }}  \label{(58)}
\end{equation}%
so that%
\begin{equation}
D\ f\left( abc\right) =0+circ.perm.  \label{(59)}
\end{equation}

\bigskip

We proved that a solution to \ref{(21)} is 
\begin{eqnarray}
\Phi \left( x_{i},\ y_{i}\right) &=&\left[ \sum_{k=0}^{\infty\,\, {\rm or\, }%
\beta -1}\frac{\Gamma \left( \beta -k\right) }{\Gamma \left( \beta +k\right) 
}\frac{1}{8^{k}\ k!}\ \ \frac{\left\{ k,k,k\right\} }{\left( abc\right)
^{2\beta }}\right] \ e^{\sum_{i=1}^{3}x_{i}y_{i}}  \label{(60a)} \\
H_{\rm Calogero}\Phi \left( x_{i},\ y_{i}\right) &=&0  \label{(60b)}
\end{eqnarray}%
where $H_{\rm Calogero}$ is given in \ref{(20)}.

\bigskip\

\end{document}